\documentclass[letterpaper,11pt]{article}

\usepackage{amsmath,amssymb,amsfonts}
\usepackage{bm,bbm}
\usepackage{epsfig}
\usepackage{cite}
\usepackage{color}

\numberwithin{equation}{section}

\newcommand{\beq}{\begin{eqnarray}}
\newcommand{\eeq}{\end{eqnarray}}
\newcommand{\nn}{\nonumber}

\newcommand{\eql}[1]{\label{eq:#1}}
\newcommand{\eq}[1]{(\ref{eq:#1})}
\newcommand{\Secl}[1]{\label{sec:#1}}
\newcommand{\Sec}[1]{Sec.\ \ref{sec:#1}}
\newcommand{\figl}[1]{\label{fig:#1}}
\newcommand{\fig}[1]{Fig.\ \ref{fig:#1}}

\newcommand{\gsim}{\mathrel{\lower.8ex\vbox{\lineskip=.1ex\baselineskip=0ex
                   \hbox{$>$}\hbox{$\sim$}}}}
\newcommand{\lsim}{\mathrel{\lower.8ex\vbox{\lineskip=.1ex\baselineskip=0ex
                   \hbox{$<$}\hbox{$\sim$}}}}

\newcommand{\too}{\longrightarrow}

\newcommand{\lan}{\langle}
\newcommand{\ran}{\rangle}
\newcommand{\lt}{\left}
\newcommand{\rt}{\right}

\DeclareMathOperator{\tr}{tr}

\newcommand{\fr}[2]{\frac{#1}{#2}}

\newcommand{\del}{\partial}

\newcommand{\td}[1]{{\tilde{#1}}}
\newcommand{\wba}[1]{{\overline{#1}}}
\newcommand{\Sla}[1]{\kern0.12em{\raise.15ex\hbox{$/$}\kern-.74em #1}}
\newcommand{\conj}{{\text{c}}}

\newcommand{\al}{\alpha}
\newcommand{\ga}{\gamma}
\newcommand{\Ga}{\Gamma}
\newcommand{\de}{\delta}
\newcommand{\De}{\Delta}
\newcommand{\ep}{\epsilon}
\newcommand{\vep}{\varepsilon}
\newcommand{\La}{\Lambda}
\newcommand{\sg}{\sigma}
\newcommand{\tht}{\theta}

\newcommand{\sgb}{\bar{\sigma}}
\newcommand{\Psb}{\wba{\Psi}}

\newcommand{\cL}{\mathcal{L}}
\newcommand{\cO}{\mathcal{O}}

\newcommand{\mum}{\text{$\mu$m}}
\newcommand{\mm}{\text{mm}}
\newcommand{\m}{\text{m}}

\newcommand{\Mev}{\text{MeV}}
\newcommand{\Gev}{\text{GeV}}
\newcommand{\Tev}{\text{TeV}}

\newcommand{\fb}{\text{fb}}

\newcommand{\SU}{\text{SU}}
\newcommand{\SO}{\text{SO}}
\newcommand{\U}{\text{U}}
\newcommand{\SP}{\text{Sp}}
\newcommand{\Z}{{\mathbb Z}}
\newcommand{\T}{{\rm T}}

\newcommand{\hrho}{{\tilde{\rho}}}
\newcommand{\hpi}{{\tilde{\pi}}}
\newcommand{\hK}{{\widetilde{K}}}

\begin{document}

%%%%%%%%%%%%%%%%
%%%%%%%%%%%%%%%%

\begin{titlepage}

\setcounter{page}{0}

\begin{flushright}
UMD-PP-09-037
\end{flushright}

\vskip 1in

\begin{center}

{\Huge\bf Vectorlike Confinement at the LHC}

\vskip .5in

{\Large {\bf Can Kilic}$^{a,1}$,
        {\bf Takemichi Okui}$^{a,b,2}$,
    and {\bf Raman Sundrum}$^{a,3}$}

\vskip .5in

$^a$~{\it Department of Physics and Astronomy, Johns Hopkins University,\\
          Baltimore, MD 21218, USA}\\
$^b$~{\it Department of Physics, University of Maryland,\\
          College Park, MD 20742, USA}

\vskip 0.5in

\abstract{We argue for the plausibility of a broad class of vectorlike
confining gauge theories at the TeV scale which interact with the
Standard Model predominantly via gauge interactions. These theories have
a rich phenomenology at the LHC if confinement occurs at the TeV scale,
while ensuring negligible impact on precision electroweak and flavor
observables. Spin-1 bound states can be resonantly produced via their
mixing with Standard Model gauge bosons. The resonances promptly decay
to pseudo-Goldstone bosons, some of which promptly decay to a pair of
Standard Model gauge bosons, while others are charged and stable on
collider time scales. The diverse set of final states with little
background include multiple photons and leptons, missing energy, massive
stable charged particles and the possibility of highly displaced
vertices in dilepton, leptoquark or diquark decays. Among others, a
novel experimental signature of resonance reconstruction out of massive
stable charged particles is highlighted. Some of the long-lived states
also constitute Dark Matter candidates.}

\end{center}

\vfill
\begin{flushleft}
$^1$~{\tt kilic@pha.jhu.edu}\\
$^2$~{\tt okui@pha.jhu.edu}\\
$^3$~{\tt sundrum@pha.jhu.edu}
\end{flushleft}

\end{titlepage}

\tableofcontents

%%%%%%%%%%%%%%%%%%%%%%%%%%%%%%%%
%%%%%%%%%%%%%%%%%%%%%%%%%%%%%%%%
\section{Introduction}
\Secl{intro}

Expectations for new physics at the LHC have been greatly influenced by
the Hierarchy Problem of electroweak symmetry breaking. However, all
proposed resolutions of this problem are strongly constrained by
existing data, either through the many direct searches for new physics
or through virtual sensitivity to new physics in a variety of precision
tests. This has led to three operational possibilities: (i) There exists
a fully satisfactory solution to the Hierarchy Problem which naturally
evades the host of experimental constraints, although we have not
discovered it yet. (ii) Naturalness, as usually interpreted, is not a
useful or valid principle of particle physics. (iii) Naturalness is an
important principle in particle physics, but it is not perfectly
satisfied in Nature, there being some modest tuning in the electroweak
symmetry breaking sector. That is, the new physics associated with
cutting off quadratic divergences in the Standard Model (SM) is heavy
enough to naturally evade existing experimental constraints, but
possibly at the cost of being out of reach of the LHC!

In either of cases (ii) or (iii), the LHC may still discover new
physics, but not directly related to the resolution of the Hierarchy
Problem. In case (ii), this would require a coincidence that the new
physics just happens to lie within the decade in energy improvement of
the LHC over the Tevatron. In case (iii), the odds are better: there may
well be some LHC-accessible ``low-energy'' tail of the spectrum of new
physics that cuts off quadratic divergences, the central parts of the
spectrum however being somewhat above the reach of the LHC. In cases
(ii) and (iii), considerations other than the hierarchy problem are
needed to constrain the form of LHC physics to anticipate. Split
Supersymmetry \cite{ArkaniHamed:2004fb} is a well known example of
either (ii) or (iii), depending on whether the scalar sparticles are
orders of magnitude heavier than the TeV scale or just modestly heavier,
respectively.

The direction taken in this paper is motivated especially by possibility (iii). From this viewpoint, one of the gauge-field-theoretically simplest and
most plausible extensions of the SM is obtained by adding new fermions in vectorlike representations of the SM gauge groups, with a mass scale within the
reach of the LHC. One can easily imagine such fermions as remnants of more involved physics at even higher energies. In contrast, adding new chiral
fermion representations of the SM gauge group is trickier, because it requires considerable particle content for anomaly cancellation, and because mass
generation via electroweak symmetry breaking generically leads to significant oblique corrections to precision electroweak observables. (However, see
Ref.\ \cite{4thgen} for discussion of the 4th generation model and viable regions of parameter space.) The advantage of a vectorlike new fermion, on the
other hand, is that it can have mass without coupling to electroweak symmetry breaking so that its impact on precision electroweak observables can be seen
to be harmless from the outset.

The new fermions may also interact via new gauge forces, which may be
weakly or strongly coupled. While the case of purely weakly coupled new
vectorlike fermions is plausible and interesting in its own right, its
collider phenomenology is straightforward and we do not pursue this
possibility here. Numerous models containing $Z'$s fall in this
category.

Instead, we will consider the case where, in addition to SM gauge
forces, the new fermions also feel a new strong gauge force that
confines at TeV energies, which drastically alters the phenomenology.
Calling the new gauge force ``hypercolor'',%
\footnote{The term hypercolor was originally used in
\cite{Eichten:1979ah} to describe what ultimately became known as
technicolor-style models. As that use has become extinct, we are
employing the term in a new way.}
the LHC phenomenology will be dominated by production of the confined
``hyperhadrons'' made from the vectorlike fermions. There may also exist
very weak nonrenormalizable interactions arising from unspecified
physics far above the TeV scale, which will only be experimentally
relevant at the LHC in mediating the decays of hyperhadrons that cannot
decay via hypercolor and SM interactions alone. We will call this
scenario ``Vectorlike Confinement''.

Vectorlike confinement represents a broad category of possible new
physics accessible to the LHC, which might represent our first glimpse
of an even richer structure at even higher energies (possibly tied to
the hierarchy problem). It has a number of attractive features from the
viewpoints of both theory and experiment, as well as prospects for rich
LHC phenomenology:
\begin{itemize}
\item{There is a close precedent in familiar particle physics. Particle
physics at low energies, below 100 MeV say, is dominated by QED, a
precise theory which is beautifully tested. Let us consider this as the
analogue of the entire SM tested up to the TeV scale, the
electromagnetic force being the analogue of all the SM gauge forces. We
can think of $e^+ e^-$ collisions at the GeV scale as the analogue of
turning on the LHC. On the one hand, for no very good reason (as far as
we know) the collisions will pair-produce muons, a ``new'' fermion with
vectorlike quantum numbers under the electromagnetic force. This is the
analogue of finding new vectorlike fermions at the LHC, but without any
new gauge forces involved at these energies. As mentioned above, this is
a straightforward possibility, but not what we will focus on.

On the other hand, we can also pair-produce the first generation quarks,
also vectorlike with respect to electromagnetism, but feeling an
additional strong gauge force, QCD. Quark confinement dominates the
phenomenology, quark pair creation being replaced by pion pair
production as well as the $\rho$ resonance. The fundamental scale is
provided by the confinement scale $\sim$ hundreds of MeV.%
\footnote{The pions, since they are light pseudo-Goldstone bosons, have
masses which are sensitive to the much smaller ``current'' quark
masses.}
This is the analogue of the Vectorlike Confinement scenario we are
proposing; as we will see the pseudo-Goldstone bosons in Vectorlike
Confinement will also play a significant role in LHC phenomenology.

While some of the hadrons produced in the $e^+ e^-$ collisions decay via
electromagnetic interactions alone, for example $\pi^{0} \rightarrow
\gamma\gamma$, others such as $\pi^{\pm}$ cannot decay via QCD or QED.
However, the weak interactions living at 100 GeV, well above the $e^{+}
e^{-}$ collision energies and the confinement scale, mediate decays of
some hadrons, {\it e.g.} $\pi^- \rightarrow \mu^-$+$\nu$'s, so that they
are merely long-lived rather than absolutely stable. Other hadrons, {\it
e.g.} the proton, can be essentially stable. Analogous mechanisms give
rise to a hierarchy of decay lifetimes in the Vectorlike Confinement
scenario as well.}
\item{In some sense, there is a tension between hopes for LHC physics
and all past experimental data. If there is exciting new physics at LHC
energies, how have its virtual effects remained so well hidden from the
myriad precision experiments we have already done? Vectorlike
Confinement accomplishes this in a very simple way. TeV-scale
confinement can indeed lead to a very rich spectrum of accessible
hyperhadron physics, as we shall discuss below. But because the dominant
bridge between the SM and the new physics is provided by the SM gauge
interactions, the new physics is flavor blind, which allows it to evade
the host of constraints from flavor experiments. Furthermore, the new
physics, being vectorlike, can be naturally separated from the
electroweak breaking sector (e.g. masses can be set by the confinement
scale), thereby evading the body of precision electroweak tests. (Other
classes of experimental data will be discussed below.)}
\item{While Vectorlike Confinement involves very modest additions to the
SM, as measured by either fundamental particle content or complexity of
Lagrangian, it can naturally give rise to a remarkable array of distinct
experimental behaviors, including di-gauge-boson resonances, long-lived
charged and/or colored states, SUSY look-alike spectra, WIMP dark matter
candidates, and leptoquarks. Many of these signals have appeared
previously as parts of other scenarios, but often within rather
exceptional models or models which are very tightly constrained already
by experiments. Here we show how such signals can arise in Vectorlike
Confinement models which have very few parameters and need not be tuned
to avoid exclusion. We will also see that there are some qualitatively
distinct signals that have not been discussed before, such as pair
production of collider stable particles through a resonance.}
\end{itemize}

This paper builds on our earlier proposal of new confining vectorlike
fermions charged under QCD, presented in \cite{coloron}, which was
however restricted in that the new fermions were electroweak singlets.%
\footnote{The Tevatron prospects for discovery in the multijet channel
were illustrated in \cite{coloron} and a follow-up study for the LHC was
presented in \cite{coloron_LHC}.}
Here, we generalize to cases where all SM gauge forces come into play. A
theoretically similar class of models with significantly different
phenomenology was provided in the ``Quirks'' model of \cite{quirks}.
These models contain new vectorlike fermions with TeV scale current
masses and extremely low confinement scales. Another closely related
class of models is the ``Hidden Valley'' scenario \cite{hidden}. While
in Vectorlike Confinement the SM gauge interactions provide the bridge
to the new physics, in hidden valley models this is done by new
interactions ($Z'$ exchange) since the new matter sector is neutral
under the SM gauge interactions.

The fact that Vectorlike Confinement involves a new TeV-scale confining
force may give the impression that it is a part of the Technicolor
scenario and phenomenology \cite{techni} (for reviews see
\cite{TC-reviews,Lane:2002sm,Brooijmans:2008se,Sannino:2008ha}). But in
fact there are very significant differences in motivation, theoretical
structure, and experimental implications. Since Technicolor aims at
solving the hierarchy problem immediately at the TeV scale,
technifermions are in a {\it chiral} representation of electroweak
symmetry, which is broken by the TeV confining force. This leads to an
impact on electroweak precision observables which poses a challenge for
the Technicolor program. Furthermore Technicolor must be closely tied to
flavor physics in order to generate SM fermion masses after electroweak
symmetry breaking%
\footnote{Walking Technicolor \cite{Appelquist:1986an} and Conformal
Technicolor \cite{Luty:2004ye} are scenarios which attempt to decouple
flavor physics to the far UV. For discussions of similarities and
differences between Walking Technicolor and Conformal Technicolor, see
Refs.\ \cite{Luty:2004ye}.}.
In contrast, Vectorlike Confinement does not break electroweak symmetry,
thus having negligible impact on electroweak precision observables
(electroweak symmetry can be broken by a more or less standard Higgs
sector, separated from the new physics). Moreover, Vectorlike
Confinement is flavor blind by construction as noted above. We do not
however exclude the possibility that Vectorlike Confinement can be a
remnant of some even higher scale physics that ameliorates the hierarchy
problem using strong dynamics. For comparisons with studies of
Technicolor phenomenology see
\cite{Lane:2002sm,Brooijmans:2008se,additionalTC}.

Calculationally, we do borrow the central trick of Technicolor analysis.
While we know that confinement, already realized in nature, is a rather
typical behavior for non-abelian gauge field theories with qualitatively
interesting consequences, the associated strong coupling makes it a
notoriously difficult phenomenon to study quantitatively. Technicolor
harnesses the power of strongly coupled gauge dynamics in a
theoretically reliable way by imitating QCD itself and using strong
interaction data as an ``analog computer''. We will follow essentially
the same strategy, differing only in how the new physics couples to the
SM gauge interactions.

This paper is aimed at highlighting some of the generic features of the
rich phenomenology in the Vectorlike Confinement scenario as well as
demonstrating consistency with all present experimental data, both in
the form of exclusions from direct searches as well as the
non-observation of any virtual effects. Further details of the
phenomenology as well as an outlook for experimental searches will be
presented in future work \cite{nextpaper}.

In section \ref{sec:micro} we will quantitatively describe the short
distance physics in the Vectorlike Confinement scenario. Next, we will
set up an effective Lagrangian in section \ref{sec:Lagrangian} to
describe the physics relevant for the LHC. Readers who are more
interested in the collider phenomenology than the theoretical details
can skip section \ref{sec:micro} and most of section
\ref{sec:Lagrangian} and jump directly to subsection \ref{sec:shortcut}
where we summarize the physics processes that are most relevant for the
LHC. We go on to discuss several representative models of Vectorlike
Confinement and their most salient features in section \ref{sec:pheno}.
Our conclusions will be summarized in section \ref{sec:conclusions}.

%%%%%%%%%%%%%%%%%%%%%%%%%%%%%%%%
%%%%%%%%%%%%%%%%%%%%%%%%%%%%%%%%
\section{The Microscopic Structure of Vectorlike Confinement}
\label{sec:micro}

The fundamental Lagrangian of Vectorlike Confinement can be best
described by drawing an analogy with the QED-QCD system (see
\fig{analogy}).
\begin{figure}
\includegraphics[width=\linewidth]{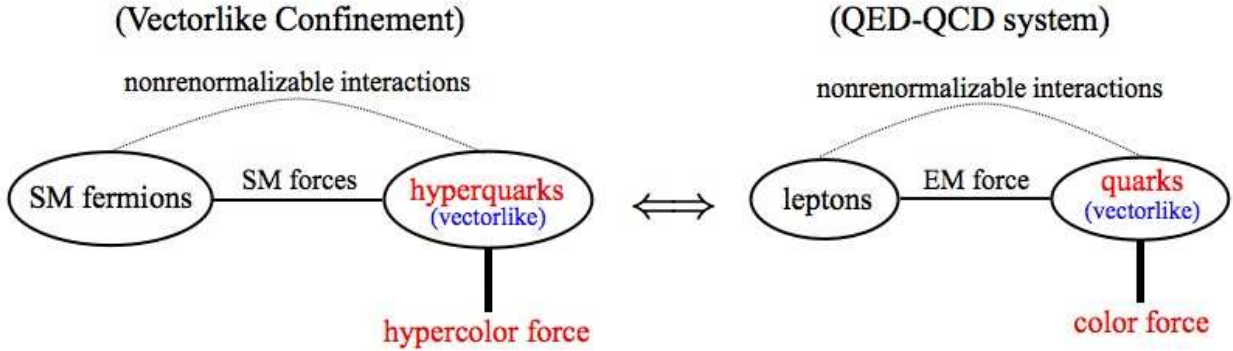}
\caption{The analogy between the structure of Vectorlike Confinement and
that of the QED-QCD system. Each oval represents a set of matter fields,
while each line represents an interaction, thicker lines corresponding
to stronger interactions. A line ending at an oval means that the oval
feels the interaction represented by the line.}
\figl{analogy}
\end{figure}
In the QED-QCD system, there are two types of matter, leptons and
vectorlike quarks. There are three kinds of interactions, color force
(strong), electromagnetic force (medium strength), and nonrenormalizable
interactions (weak) such as 4-fermion operators with two leptons and two
quarks. The nonrenormalizable interactions are quite feeble, and their
major role in phenomenology is to break otherwise conserved quantum
numbers (such as ``up number'' and ``down number'') and let otherwise
stable particles (such as $\pi^\pm$) decay. If we ignore the
nonrenormalizable interactions, the electromagnetic gauge interaction is
the only interaction connecting quarks to leptons.

Similarly, in Vectorlike Confinement, we have two sets of matter, SM
fermions and vectorlike ``hyperquarks''. Hyperquarks feel three types of
interactions, a new confining gauge interaction ``hypercolor'' (strong),
SM gauge interactions (medium strength), and nonrenormalizable
interactions (weak). The nonrenormalizable interactions have very small
effects, and their major phenomenological impact is to break otherwise
exact symmetries and let otherwise stable particles decay. If we ignore
the nonrenormalizable interactions, hyperquarks can interact with SM
fermions only via SM gauge interactions.\footnote{Ensuring this property
requires in particular that the quantum numbers of hyperquarks be such
that no Yukawa couplings with the Higgs are allowed for hyperquarks.}

The vectorlike nature of hyperquarks is crucial. Just like color, hypercolor is assumed to confine and spontaneously break the axial flavor symmetry of
hyperquarks while preserving the vector flavor symmetry. Since hyperquarks are vectorlike under SM gauge interactions, this ``hyper-chiral symmetry
breaking'' does {\it not} break any of the SM gauge symmetries. This (and the absence of renormalizable Higgs-hyperquark couplings) guarantee that the
hypercolor dynamics have negligible impact on electroweak precision observables \cite{PrecisionEW, PT}. More systematically, note that, upon integrating
out the hyper sector, the nonrenormalizable operators that parameterize corrections to precision electroweak observables (e.g.\ $(H^\dag W_{\mu\nu} H)
B^{\mu\nu}$ for the $S$ parameter) all involve $H$.  But since hyperquarks interact with the SM only via gauge interactions and have no direct couplings
to $H$, one sees that any diagram with hyperquarks that contributes to those operators must involve {\em additional} two loops beyond {\em existing} SM
contributions. Note that hyperquark loops do contribute to electroweak-preserving vacuum polarizations such as $W^{\mu\nu} W_{\mu\nu}$ and $B^{\mu\nu}
B_{\mu\nu}$, but their effects on precision electroweak observables cancel out once expressed in terms of the standard input parameters such as the fine
structure constant, $G_F$ and $m_Z$, as is clear from the Peskin-Takeuchi analysis \cite{PT}.

Another crucial feature depicted by \fig{analogy} is that, in the limit
of neglecting nonrenormalizable interactions, SM gauge interactions are
the only interactions connecting hyperquarks to SM fermions. Then, in
this limit, flavor violations beyond the SM are completely negligible,
since new flavor-violating diagrams involving hyperquark loops are
necessarily suppressed by additional two loop factors ($\sim
1/(16\pi^2)^2$) compared to the {\it existing} flavor-violating diagrams
in the SM. Nonrenormalizable operators, especially 4-fermion operators
with two or four SM fermions, do in general induce new flavor
violations, but they {\it can} be suppressed by simply taking the scale
of these operators to be high.

Below, we will describe the theory of the Vectorlike Confinement
scenario and analyze its dynamics by beginning with the strongest
interaction, hypercolor, and then turning on weaker interactions, one by
one, as perturbations.

%%%%%%%%%%%%%%%%%%%%%%%%%%%%%%%%
%%%%%%%%%%%%%%%%%%%%%%%%%%%%%%%%
\subsection{Hypercolor and Hyperflavor}

Let us begin with the strongest interaction, i.e.\ hypercolor (HC), and
ignore all SM interactions and nonrenormalizable interactions. The
Lagrangian in this limit reads simply
\beq
  \cL
  = \cL_\text{SM}
    -\frac{1}{4} H^A_{\mu\nu} H^{A\mu\nu}
    +\sum_{i=1}^F \overline{\psi}_i (i\Sla{D} - m_i) \psi_i  \,,
\eql{pure-hyper}
\eeq
where $H^A_{\mu\nu}$ is the field strength for the $\SU(N)$ hypercolor
gauge field $H^A_\mu$ ($A=1$, $\cdots$, $N^2-1$), and
\beq
  D_\mu = \del_\mu + ig_\text{HC} H^A_\mu t^A \,,
\eeq
where $g_\text{HC}$ is the hypercolor gauge coupling and $t^A$ is a
generator of $\SU(N)$. There are $F$ Dirac fermions $\psi_i$ ($i=1$,
$\cdots$, $F$), or ``hyperquarks'', all of which are assumed to be in
the fundamental representation of $\SU(N)_\text{HC}$.\footnote{If we
choose a non-complex representation of $\SU(N)_\text{HC}$ for
hyperquarks, or if we choose a non-complex group like $\SO(N)$ or
$\SP(N)$ for hypercolor, then the hyperflavor breaking pattern
\eq{hyper-chiral-symm-break} would be changed and the analogy with the
QED-QCD system must be used with care. We will not pursue such
alternatives in this paper.} We refer to the variety of $\psi_i$
distinguished by $i$ as ``hyperflavor'', analogous to flavor in QCD. If
we neglect hyperquark masses $m_i$, the theory has a global symmetry
(``hyperflavor symmetry'')
\beq
  \SU(F)_\text{L} \otimes \SU(F)_\text{R} \otimes \U(1)_\text{HB}
\eql{hyper-flavor}
\eeq
under which $\psi_\text{L,R} \equiv \fr{1 \mp \ga_5}{2} \psi$ (where
$\psi \equiv (\psi_1, \cdots, \psi_F)$) transform as
\beq
  \psi_\text{L} &\too& \exp[i\al_\text{L}^a T^a] \, \psi_\text{L}  \,,\nn\\
  \psi_\text{R} &\too& \exp[i\al_\text{R}^a T^a] \, \psi_\text{R}  \,,\nn\\
  \psi          &\too& \exp[i\al_\text{HB}/N] \, \psi  \,,
\eql{hyper-flavor-transformation}
\eeq
where $T^a$ ($a=1, \cdots, F^2-1$) are the generators of $\SU(F)$.
$\U(1)_\text{HB}$, or ``hyperbaryon number'', is analogous to the baryon
number in QCD.

Like QCD, we assume that hypercolor confines and triggers spontaneous
breaking of the hyperflavor symmetry \eq{hyper-flavor} as:
\beq
  \SU(F)_\text{L} \otimes \SU(F)_\text{R} \otimes \U(1)_\text{HB}
  &\too& \SU(F)_\text{V} \otimes \U(1)_\text{HB}
\eql{hyper-chiral-symm-break}
\eeq
where $\SU(F)_\text{V}$ is the subgroup satisfying $\al_\text{L} =
\al_\text{R}$ in \eq{hyper-flavor-transformation}. Nonzero hyperquark
masses $m_i$ do not significantly impact this symmetry breaking pattern
as long as they are much smaller than the hypercolor confinement scale
$\La_\text{HC}$. (Therefore, we {\it define} $F$ to be the number of
light ($\ll \La_\text{HC}$) hyperquarks.) Note that the value of $F$
must be consistent (i.e.\ sufficiently small) with the occurrence of
hypercolor confinement and ``hyper-chiral symmetry breaking''
\eq{hyper-chiral-symm-break}.

Analogously to QCD, hyper-chiral symmetry breaking yields $F^2-1$
(Pseudo)-Nambu-Goldstone bosons transforming altogether as an adjoint
multiplet of $\SU(F)_\text{V}$ (with small masses induced by $m_{i}$).
We refer to them as ``hyperpions''. Being much lighter than all other
bound states of hyperquarks, hyperpions play an important role in
collider phenomenology.

%%%%%%%%%%%%%%%%%%%%%%%%%%%%%%%%
%%%%%%%%%%%%%%%%%%%%%%%%%%%%%%%%
\subsection{SM Gauge Interactions and the Concept of ``Species''}
\Secl{SM-gauge}

Let us now turn on SM gauge interactions. The fundamental Lagrangian now reads
\beq
  \cL
  = \cL_\text{SM}
    -\frac{1}{4} H^A_{\mu\nu} H^{A\mu\nu}
    +\sum_{I=1}^{S} \overline{\Psi}_I (i\Sla{D} - m_I) \Psi_I  \,,
\eql{addingSMgauge}
\eeq
where
\beq
  D_{\mu} \Psi_I
  = \del_\mu \Psi_I + ig_\text{HC} H^A_\mu t^A \Psi_I
                      + i\sum_G g_G A^{(G)\al}_{\mu} T_{\Psi_I}^{(G)\al} \Psi_I \,,
\eeq
where $T_{\Psi_I}^{(G)}$ are the generators of the SM gauge group $G =
\SU(3)_\text{C}$, $\SU(2)_\text{W}$, $\U(1)_\text{Y}$ for $\Psi_I$. Note
that $\Psi_I$ is {\it different} from $\psi_i$ in \eq{pure-hyper}; the
$F$ flavors, $\psi_1$, $\psi_2$, $\cdots$, $\psi_F$, have reorganized
themselves into $S$ multiplets, $\Psi_1$, $\Psi_2$, $\cdots$,
$\Psi_{S}$, of the SM gauge group. We refer to the variety of $\Psi_I$
distinguished by $I$ as ``species''. For example, one could imagine a
5-flavor ($F=5$) model where the 5 flavors reorganize themselves into
two species ($S=2$), say, a triplet of color and a doublet of weak
interaction.%
\footnote{The distinction between species and flavors is obscured in the
QED-QCD analogy. In this analogy, $G = \U(1)_\text{EM}$ and there are
three species $u$, $d$, $s$ with $T_{u} = 2/3$, $T_{d} = -1/3$, $T_{s} =
-1/3$. But since $\U(1)$ has only one-dimensional representations,
species coincide with flavors.}
Now that the $\Psi$ are charged under the SM, one must make sure that their impact on the running of SM gauge couplings is small enough to avoid Landau
poles near the TeV scale.

By our assumption the $\Psi$ are all vectorlike under the SM gauge
group, that is, $T_{\Psi_I}^{(G)}$ all belong to $\SU(F)_\text{V}$.
Therefore, SM gauge interactions explicitly break the axial part of
$\SU(F)_\text{L} \otimes \SU(F)_\text{R}$, so hyperpions that carry SM
charges are not exact Nambu-Goldstone bosons even if $m_{I}=0$.
Consequently, non-singlet hyperpions acquire mass contributions
depending on their SM gauge quantum numbers. This is analogous to the
difference $m_{\pi^\pm}^2 - m_{\pi^0}^2$ in the QED-QCD system, which is
induced by QED. Note that the radiative contributions from the SM gauge
sector can be the dominant source of mass for hyperpions, whereas in QCD
the up and down quark masses are large enough to be the dominant source
of pion masses.

%%%%%%%%%%%%%%%%%%%%%%%%%%%%%%%%
%%%%%%%%%%%%%%%%%%%%%%%%%%%%%%%%
\subsubsection{Mixing of Vector Mesons with SM Gauge Bosons}
\Secl{mixings}

The perturbation added in going from \eq{pure-hyper} to \eq{addingSMgauge} is
\beq
  \De\cL
  = -\sum_G g_G A^{(G)\al}_{\mu} J_G^{\al\mu}  \,,
\eql{A-rho-mix}
\eeq
where $G = \SU(3)_\text{C}$, $\SU(2)_\text{W}$, $\U(1)_\text{Y}$, and
\beq
   J_G^{\al\mu}
   = \sum_{I=1}^S \overline{\Psi}_I \ga^\mu T_{\Psi_I}^{(G)\al} \Psi_I  \,.
\eeq
Now, analogously to $\rho$ mesons in QCD, the Noether currents
corresponding to $\SU(F)_\text{V} \otimes \U(1)_\text{HB}$ can create
parity-odd spin-1 bound states of a hyperquark and an anti-hyperquark,
or ``hyper-$\rho$ mesons'' for short, with masses $m_\hrho \sim
\La_\text{HC}$. Since $J_G^{\al\mu}$ are a subset of these vector
currents, the interactions \eq{A-rho-mix} lead to mixings between SM
gauge bosons and the hyper-$\rho$ mesons interpolated by $J_G^\mu$. In
the QED-QCD system, such a mixing occurs between the $\ga$, $\rho^0$ and
$\omega$.

The hyper-$\rho$ mesons corresponding to $J_G^{\al\mu}$ have a special
status, as they can be {\it singly} produced via mixing with an
$s$-channel SM gauge boson, while the rest of hyper-$\rho$'s and all
other massive hyper-hadrons must be pair-produced. Therefore, together
with the hyperpions, which are light, these hyper-$\rho$'s dominate the
collider phenomenology.

Let us revisit the corrections to precision electroweak observables and see why they are kept small in Vectorlike Confinement, this time from the
``hadronic'' viewpoint of $\hrho$-SM-gauge-boson mixing. New physics contributions to the ``oblique'' parameters, i.e.\ vacuum polarizations of
electroweak gauge bosons, are dominated by the mixing of electroweak gauge bosons to the $\hrho$ and back again. Since the $\hrho$ are much heavier than
the $Z$ mass, the oblique corrections are well parameterized by the Peskin-Takeuchi $S$, $T$ and $U$ parameters \cite{PT}. Since $T$ and $U$ measure
custodial isospin violation and the hyper sector explicitly preserves this symmetry, we need only be concerned with the $S$ parameter. $S$ is essentially
the strength of kinetic mixing $W_{\mu \nu}^3 B^{\mu \nu}$ after electroweak symmetry breaking, which feeds into precision observables such as the $W$
mass measurement. However in our case, the new physics does not break electroweak symmetry itself, and neither does it couple to the Higgs, so mixing with
the $\hrho$ only induces the {\em electroweak-preserving} polarization effects, $W_{\mu \nu} W^{\mu \nu}$ and $B_{\mu \nu} B^{\mu \nu}$, which however do
not feed into precision electroweak observables as we already discussed at the beginning of \Sec{micro}.

%%%%%%%%%%%%%%%%%%%%%%%%%%%%%%%%
%%%%%%%%%%%%%%%%%%%%%%%%%%%%%%%%
\subsubsection{Accidental ``Species Symmetries'' and Long-lived Hyperpions}
\Secl{accidental}

In the renormalizable Lagrangian \eq{addingSMgauge}, the SM gauge
interactions explicitly break $\SU(F)_\text{V} \otimes \U(1)_\text{HB}$
flavor symmetry, but the subgroup $\U(1)_{\Psi_1} \otimes \U(1)_{\Psi_2}
\otimes \cdots \otimes \U(1)_{\Psi_S}$ are still preserved. Therefore,
``$\Psi_1$ number'', ``$\Psi_2$ number'', $\cdots$, ``$\Psi_S$ number'',
or collectively ``species numbers'', are all conserved.\footnote{A
species symmetry could in principle be larger than $\U(1)$ if there are
more than one species with identical SM gauge quantum numbers and
masses. For example, if $\Psi_1$ and $\Psi_2$ have the same gauge
quantum numbers and masses, they would have a $\U(2)$ species symmetry,
instead of $\U(1)_{\Psi_1} \otimes \U(1)_{\Psi_2}$. However, unlike
$\U(1)$ species symmetries, which are automatic in the renormalizable
Lagrangian \eq{addingSMgauge}, a non-Abelian species symmetry requires
tuning of masses and therefore does not enjoy the privilege of {\it
accidental} symmetry. While one could remove the tuning by gauging the
species symmetry, which may give rise to interesting phenomenology, we
will not pursue this possibility further in this paper.} A special case
is the hyperbaryon number, which is just the average of all species
numbers.

Species number conservation has profound phenomenological consequences.
A hyperpion with the quantum numbers of $\overline{\Psi}_I \Psi_J$ ($I
\neq J$) cannot decay to SM particles due to its nonzero species
numbers. In the QED-QCD analogy of \fig{analogy}, the $\pi^\pm$ would be
exactly stable if the nonrenormalizable interactions were ignored, due
to the conservation of ``up number'' and ``down number''. On the other
hand, a hyperpion of the $\wba{\Psi}_I \Psi_I$ type carries no species
numbers and will promptly decay to a pair of SM gauge bosons, just like
$\pi^0 \to \ga\ga$ in the QED-QCD system.

Of course, species number conservation is an accidental feature of the
renormalizable Lagrangian \eq{addingSMgauge}, so these stable particles
may eventually decay via a nonrenormalizable interaction, just like the
$\pi^\pm$ eventually decays to leptons via a 4-fermion operator. This is
the topic of the following subsection.

%%%%%%%%%%%%%%%%%%%%%%%%%%%%%%%%
%%%%%%%%%%%%%%%%%%%%%%%%%%%%%%%%
\subsection{Nonrenormalizable Interactions}

We finally come to the last ingredient depicted in \fig{analogy}, namely
nonrenormalizable operators connecting the SM and the hyper sector.
Nonrenormalizable interactions are of course expected to exist at some
level, and while most of them do not qualitatively affect the physics,
some of them do as they induce processes forbidden in the renormalizable
Lagrangian \eq{addingSMgauge}, such as those which violate species
numbers.

%%%%%%%%%%%%%%%%%%%%%%%%%%%%%%%%
%%%%%%%%%%%%%%%%%%%%%%%%%%%%%%%%
\subsubsection{Mixed SM-hyper 4-fermion Interactions and Hyperpion Decays}
\Secl{4-fermion}

In the QED-QCD system, the 4-fermion operators $(\bar{d}\ga_\mu \ga_5 u)
(\bar{\nu}_\ell \sgb^\mu \ell)$ with $\ell = e, \mu$ are the leading
interactions which break the up and down numbers and allow a $\pi^+$ to
decay. Similarly, the lifetimes of hyperpions with nonzero species
numbers are controlled by the 4-fermion interactions involving two
hyperquarks and two SM fermions which break species
symmetries.\footnote{It is in principle possible that a particular model
may not allow any such 4-fermion operators that are gauge invariant and
one would have to consider even higher dimensional operators. For
definiteness, we stick to the 4-fermion case in this paper.}

Suppose the $\hpi$ with the quantum numbers of $\wba{\Psi}_I \Psi_J$
decays to a pair of SM fermions, either $\bar{f}_i f'_j$ or $f_i f'_j$
($f,f'=q$, $u^c$, $d^c$, $\ell$, $e^c$ and $i,j = 1,2,3$), whichever the
quantum numbers match. The 4-fermion operator mediating the
$\hpi_{\bar{I}J} \to \bar{f}_i f'_j$ decay takes the ``current-current''
form:
\beq
  \fr{C_{\hpi_{\bar{J}I} \bar{f}_i f_j}}{M^2} (\wba{\Psi}_J \ga^\mu \ga_5 \Psi_I)
  (\bar{f}_i \sgb_\mu f'_j)  \,,
\eql{current-current}
\eeq
while the one for $\hpi_{\bar{I}J} \to f_i f'_j$ takes the
``scalar-scalar'' form:
\beq
  \fr{C_{\hpi_{\bar{J}I} f_i f_j}}{M^2} (\wba{\Psi}_J \ga_5 \Psi_I) (f_i f'_j)  \,.
\eql{scalar-scalar}
\eeq
Recall that in the QED-QCD analogy, the $\pi^\pm$ decays to leptons are
induced by the current-current 4-fermion operators.%
\footnote{Of course, in the QED-QCD analogy the physics underlying the 4-fermion weak interactions ultimately is related to the origin of quark masses. In
our case, we will stay agnostic as to whether our 4-fermion operators are deeply connected to the origin of hyperfermion masses or not. For the
phenomenology we will pursue, what matters is that they in general can break accidentally conserved global quantum numbers of the renormalizable
dynamics.}

The scale suppressing these mixed SM-hyper 4-fermion interactions cannot
be too low. Using any one of these vertices twice and closing the
hyperquarks in a loop will induce a purely-SM 4-fermion operator. Unless
we make some extra assumptions governing the flavor structures of
$C_{\hpi_{\bar{J}I} \bar{f}_i f_j}$ and $C_{\hpi_{\bar{J}I} f_i f_j}$,
the induced purely-SM 4-fermion operators generically lead to
flavor-changing neutral currents (FCNCs). The absence of FCNCs then
requires the coefficients of these operators be $\lsim
(\cO(10^5)\>\Tev)^{-2}$ \cite{UTfit}. For $C_{\hpi_{\bar{J}I} \bar{f}_i
f_j} \sim C_{\hpi_{\bar{J}I} f_i f_j} \sim \cO(1)$, this translates to
the bound $M \gsim \cO(10^5/4\pi)$ TeV $\sim \cO(10^4)$ TeV. If we do
assume an extra flavor symmetry, such as $C_{\hpi_{\bar{J}I} \bar{f}_i
f_j} \propto C_{\hpi_{\bar{J}I} f_i f_j} \propto \de_{ij}$, then $M$
needs to be only high enough for the effective field theory description
of \fig{analogy} to make sense.

%%%%%%%%%%%%%%%%%%%%%%%%%%%%%%%%
%%%%%%%%%%%%%%%%%%%%%%%%%%%%%%%%
\subsubsection{Hyperbaryon Decay}

Hyperbaryon number $\U(1)_\text{HB}$ is also an accidental symmetry of
the renormalizable lagrangian \eq{addingSMgauge}. While all hypermesons
are neutral under $\U(1)_\text{HB}$, ``hyperbaryon'' operators, i.e.\
the products of $N$ hyperquark fields, carry a unit hyperbaryon number.
Since the properties of hyperbaryons strongly depend on the details such
as $N$ and $F$, we will discuss their decays in each specific model
below.

%%%%%%%%%%%%%%%%%%%%%%%%%%%%%%%%
%%%%%%%%%%%%%%%%%%%%%%%%%%%%%%%%
\section{Phenomenological Lagrangian for the LHC}
\label{sec:Lagrangian}

In this section, we write down a crude Lagrangian describing the
interactions of hyperpions ($\hpi$) and hyper-$\rho$ mesons ($\hrho$)
with the SM particles and among themselves. This description is good
enough for a rough understanding of the LHC phenomenology of the theory
introduced in \Sec{micro}. It is {\it not} meant to be a systematic
effective Lagrangian controlled by small expansion parameters. There is
some rationale for including the $\hrho$ but not other massive
hyper-hadrons, even though their masses are similar to those of the
$\hrho$. As we already pointed out in \Sec{mixings}, a $\hrho$ can be
singly produced in the collision of SM particles, while all other
massive hyperhadrons must be pair-produced, so their mass scale is
effectively $\approx 2m_\hrho$ as far as collider phenomenology is
concerned. Therefore, at hadron colliders, the signatures of the hyper
sector are dominated by the resonant production of a $\hrho$ and its
subsequent decay to a $\hpi$ pair, as well as the ``Drell-Yan''
production of a $\hpi$ pair from SM gauge interactions. In this spirit,
there is no need to include the $\hrho$ which do not mix with SM gauge
bosons as they can only be pair-produced, so our phenomenological
Lagrangian captures all interactions that are relevant for the
production and decay of the hypercolor bound states that can be produced
at energies below $\Lambda_{HC}$.

In the following, we will go through the various features of the
following phenomenological Lagrangian:
\begin{subequations}
\eql{pheno-lag}
\beq
  \cL_\text{eff}
  &=& \cL_\text{SM}  \nn\\
  && +\sum_G \lt[ -\frac14 \hrho^{(G)\al}_{\mu\nu} \hrho^{(G)\al\mu\nu}
                  +\frac{m_\hrho^2}{2} \hrho^{(G)\al}_{\mu} \hrho^{(G)\al\mu}
                  -\frac{\vep_G}{2} \hrho^{(G)\al}_{\mu\nu} F^{(G)\al\mu\nu}
             \rt]  \eql{rho-gauge-mix}\\
  && +\sum_{P=1}^{S_\hpi} \fr{1}{2^{c_P}}
      \lt[ (D_\mu \hpi_P)^\dag (D^\mu \hpi_P) -m_{\hpi_P}^2 \hpi_P^\dag \hpi_P \rt]
         \eql{pi-kinetic-mass}\\
  && -g_{\hrho\hpi\hpi} f^{abc} \hrho^a_\mu \hpi^b D^\mu \hpi^c  \eql{rho-pi-pi}\\
  && -\fr{N \ep^{\mu\nu\rho\sg}}{16\pi^2 f_\hpi}
      \sum_{G,G'} g_G g_{G'}
      \tr\bigl[ \hpi^{a}T^{a} F^{(G)}_{\mu\nu} F^{(G')}_{\rho\sg} \bigr]  \eql{pi-F-F}\\
  && +\cL_{\hpi f \bar{f}}  \,.\nn
\eeq
\end{subequations}

\subsection{Production and Decay of $\hrho$}

The first and second terms in \eq{rho-gauge-mix} are respectively the
kinetic and mass terms of $\hrho^{(G)\al}_\mu$, where
$G=\SU(3)_\text{C}$, $\SU(2)_\text{W}$, $\U(1)_\text{Y}$ and $\al$
labels the generators of each $G$. The third term describes the mixing
of $\hrho^{(G)}_\mu$ with the SM gauge boson $A^{(G)}_\mu$ (whose field
strength is $F^{(G)}_{\mu\nu}$). The precise value of the mixing
parameter $\vep_G$ depends on the details of the hyper sector, but a
rough estimate based on standard large-$N$ counting is:
\beq
  \vep_G \approx \fr{\sqrt{N}}{4\pi} g_G  \,.
\eql{mixing}
\eeq
For practical calculations it is more convenient to eliminate the mixing
term in \eq{rho-gauge-mix} and work in the ``diagonal'' basis of
$A^{(G)}_\mu$ and $\hrho^{(G)}_\mu$. This can be done (up to
$\cO(\vep_G^2)$ corrections) by shifting $A^{(G)}_\mu$ as
\beq
  A^{(G)}_\mu \too A^{(G)}_\mu - \vep_G \hrho^{(G)}_\mu  \,.
\eeq
These shifts induce couplings of the $\hrho^{(G)}_\mu$ to SM fermions
$f=q$, $u^c$, $d^c$, $\ell$, $e^c$:
\beq
  \cL_{\hrho f\bar{f}}
  = \sum_f \sum_G \vep_G g_G \bar{f} \sgb^\mu \hrho^{(G)}_\mu f  \,.
\eql{rho-f-f}
\eeq
We briefly remark here that while \eq{rho-f-f} allows resonant
production of a $\hrho$ from a $q$-$\overline{q}$ initial state, gauge
invariance does not allow for a colored $\hrho$ to be resonantly
produced from a gluon-gluon initial state via renormalizable couplings.
There are however operators of higher mass dimension that can induce
such a coupling, which are not included in \eq{pheno-lag}. In
\cite{coloron_LHC}, such an operator was studied in order to estimate
any additional contribution to the production cross section. The impact
on the phenomenology was found to be a minor one.

Once the $\hrho$ are resonantly produced, they will decay dominantly to
a pair of $\hpi$ as described by the interaction \eq{rho-pi-pi}, where
$a,b,c = 1, \cdots, F^2-1$ and $f^{abc}$ is the structure constant of
$\SU(F)_\text{V}$. The $\hrho \to \hpi\hpi$ rate (summed over all
$\hpi$) can readily be computed as
\beq
  \Ga_{\hrho \to \hpi \hpi}
  = \fr{m_\hrho g_{\hrho\hpi\hpi}^2 F}{96\pi}
    \biggl( 1- \frac{4m_\hpi^2}{m_\hrho^2} \biggr)^{\! 3/2} ,
\eql{rho-width} \eeq
where we have neglected differences among $m_{\hpi}$'s for simplicity.
The coupling strength $g_{\hrho\hpi\hpi}$ can be estimated as
\beq
  g_{\hrho\hpi\hpi} \approx \fr{4\pi}{\sqrt{N}}  \,.
\eeq
Using this estimate with $N=3$, \eq{rho-width} agrees well with the
observed $\rho \to \pi\pi$ rate in the QED-QCD system.

The $\hrho$ can also decay to SM fermions via \eq{rho-f-f}. Neglecting
the $f$ mass, the partial width {\it per Weyl fermion $f$} is given by
\beq
  \Ga_{\hrho^{(G)} \to f\bar{f}}
  = \fr{g_G^2 \vep_G^2}{24\pi} D_f m_\hrho  \,,
\eql{rho-width-to-ff}
\eeq
where $D_f = 1/2$ for $G=\SU(3)_\text{C}$, $\SU(2)_\text{W}$ while $D_f
= Y_f^2$ for $G=\U(1)_\text{Y}$. (Estimating $\vep_G$ from \eq{mixing}
with $N=3$ and $G=\U(1)_\text{EM}$, the above formula agrees well with
the partial width of $\rho^0 \to e^+ e^-$ in the QED-QCD system.) Note
that unless $N$ is very large, the decay into SM fermions is very
suppressed and the $\hrho$ width is well approximated by \eq{rho-width}.

\subsection{Sources of Hyperpion Masses}

Next, \eq{pi-kinetic-mass} describes the kinetic and mass terms of
$\hpi_P$, where $P=1$, $\cdots$, $S_\hpi$ labels different species of
$\hpi$. (When the $F$ $\psi$'s reorganize themselves into the $S$
$\Psi$'s, the $F^2-1$ $\hpi$'s accordingly reorganize themselves into
$S_\hpi$ multiplets of the SM gauge group.) The number $c_P$ is 0 for a
complex $\hpi_P$ and 1 for a real $\hpi_P$. The SM gauge interactions of
$\hpi_P$ are encoded in
\beq
  D_\mu\hpi_P
  = \del_\mu \hpi_P + i\sum_G g_G A^{(G)\al}_\mu T_{\hpi_P}^{(G)\al} \hpi_P  \,,
\eeq
where $T_{\hpi_P}^{(G)}$ are the generators of the SM gauge group $G$
for $\hpi_P$. There are three sources of $\hpi$ masses. First, as we
mentioned in \Sec{SM-gauge}, the $\hpi$ are not exact Nambu-Goldstone
bosons even when all hyperquark masses $m_I$ are zero in
\eq{addingSMgauge}, because SM gauge interactions explicitly break the
axial part of $\SU(F)_\text{L} \otimes \SU(F)_\text{R}$. These
contributions are given by
\beq
  \lt. m_{\hpi_P}^2 \rt|_\text{gauge}
  = \fr{3 c^2 m_{\hrho}^2}{16\pi^2} \sum_G g_G^2 C_2^{(G)}(\hpi_P) \,,
\label{eq:pi-mass-gauge}
\eeq
where $C_2^{(G)}(\hpi_P)$ is the quadratic Casimir of the SM gauge group
$G$ for $\hpi_P$. The constant $c$ is an $\cO(1)$ number whose precise
value depends on the details of the hyper sector. (For $N=F=3$, we
obtain $c \simeq 1.10$ from $m_{\pi^\pm}^2 - m_{\pi^0}^2$ in the QED-QCD
system.) The second source of $m_\hpi$ is the nonzero (but $\ll
m_\hrho$) hyperquark masses, giving rise to
\beq
  \lt. m_{\hpi_P}^2 \rt|_\text{h.q.\ mass}
  \approx m_{\hrho} m_\text{hq}  \,,
\eql{pi-mass-from-mhq}
\eeq
where we have taken $m_I = m_\text{hq}$ for all $I$ for simplicity. The
last possible source of $m_\hpi$ is electroweak symmetry breaking
(EWSB). If a $\hpi_P$ is charged under $\SU(2)_\text{W}$, $W$ and $Z$
loops after EWSB split the masses of different $\SU(2)_\text{W}$
components of the $\hpi_P$. The leading contributions are IR dominated
and thus calculable \cite{minimalDM}:
\beq
  \De m_\hpi
  &\simeq& \alpha_2 M_Z \, Q_\text{diff}
           \bigl[ Q_\text{avg} \sin^2\!\tht_W
                 -(Q_\text{avg} - Y)(1 - \cos\tht_W) \bigr]  ,
\eql{mass-split-IR}
\eeq
where $\De m_\hpi$ is the mass splitting between any two components of
the $\hpi_P$, and $Q_\text{diff}$ and $Q_\text{avg}$ are respectively
the difference and average of the electric charges of the two
components, and $Y$ is the common hypercharge. These mass splittings are
typically of $\cO(100)$ MeV, but can significantly affect collider
phenomenology because a heavier component can decay to a lighter
component by emitting an off-shell $W$.

\subsection{Decays of Short-lived Hyperpions}

The interactions \eq{pi-F-F} represent the chiral anomalies of the
underlying $\SU(F)_\text{L} \otimes \SU(F)_\text{R}$ in the presence of
SM gauge fields (hence the trace is taken over the $\SU(F)_\text{V}$
space, on which the SM gauge groups act). In the language of Feynman
diagrams, the triangle diagram exists only when the $\Psi$ loop can be
closed, thus these vertices exist only for the $\hpi$ made out of the
same species of $\Psi$, i.e. precisely the $\hpi$ that carry no net
species number. Such $\hpi$ will decay promptly to a pair of SM gauge
bosons. In the QED-QCD analogy, this corresponds to the well known
chiral anomaly responsible for $\pi^0 \to \ga\ga$. The $\hpi$ decay
constant $f_\hpi$ is normalized as
\beq
  \bigl\lan 0 \bigl| \Psb \ga_\mu\ga_5 T^a \Psi(x) \bigr| \hpi^b(p) \bigr\ran
  \equiv -i f_{\hpi} \de^{ab} p_\mu \, e^{-ip\cdot x}  \,,
\eql{decay-constant}
\eeq
where the $\SU(F)$ generators $T^a$ are normalized as $\tr[T^a
T^b]=\de^{ab}/2$. (This is the same as the convention where $f_\pi = 92$
MeV in the QED-QCD system.) A rough estimate for $f_\hpi$ is
\beq
  f_{\hpi} \approx \fr{m_{\hrho} \sqrt{N}}{4\pi}  \,.
\eql{f-pi}
\eeq

\subsection{Decays of Long-lived Hyperpions}

Finally, $\cL_{\hpi f\bar{f}}$ in \eq{pheno-lag} describes the
interactions which violate species symmetries, thereby allowing the
otherwise stable $\hpi$ to decay. In calculating the matrix element for
$\hpi_{\bar{I}J} \to \bar{f}_i f'_j$, the current-current operator
\eq{current-current} reduces to
\beq
  \approx \fr{C_{\hpi_{\bar{I}J} \bar{f}_i f_j}}{M^2} f_\hpi p_\hpi^\mu \,
          \hpi_{\bar{I}J}^\dag \, \bar{f}_i \sgb_\mu f'_j\label{eq:pi_decay_op_vector}
\eeq
by using the definition \eq{decay-constant}. On the other hand, for
$\hpi_{\bar{I}J} \to f_i f'_j$, the scalar-scalar operator
\eq{scalar-scalar} reduces to
\beq
  \approx \fr{C_{\hpi_{\bar{I}J} f_i f_j}}{M^2} f_\hpi m_\hrho \,
          \hpi_{\bar{I}J}^\dag \, f_i f'_j\,.\label{eq:pi_decay_op_scalar}
\eeq
Therefore, these decays can be described by including the following
effective vertices in $\cL_{\hpi f\bar{f}}$:
\beq
  \cL_{\hpi f\bar{f}}
  = \sum_{\hpi, f_i, f'_j} \fr{C_{\hpi \bar{f}_i f'_j} f_\hpi}{M^2}
                           (\del_\mu \hpi^\dag) \bar{f}_i \sgb^\mu f'_j
   +\sum_{\hpi, f_i, f'_j} \fr{C_{\hpi f_i f'_j} f_\hpi m_\hrho}{M^2}
                           \hpi^\dag f_i f'_j  \,,
\eql{L-pi-f-f}
\eeq
where $f$, $f' = q$, $u^c$, $d^c$, $\ell$, $e^c$ and $i,j = 1,2,3$. The
decay rate for $\hpi \to \bar{f}_i f'_j$ is then given (up to a group
theory factor) by
\beq
  \Ga_{\hpi \to \bar{f}_i f'_j}
  &=& \fr{C_{\hpi \bar{f}_i f'_j}^2}{16\pi} \fr{m_{f_i}^2+m_{f'_j}^2}{m_\hpi^2}
      \fr{f_\hpi^2 m_\hpi^2}{M^4} \, m_\hpi  \nn\\
  &\approx& \lt[ 10^8\>\m \times
            \fr{1}{C_{\hpi \bar{f}_i f'_j}^2 N} \fr{\Gev^2}{m_{f_i}^2+m_{f'_j}^2}
            \fr{100\,\Gev}{m_\hpi}
            \lt( \fr{\Tev}{m_\hrho} \rt)^{\!2} \!
            \lt( \fr{M}{10^4\,\Tev} \rt)^{\!4} \rt]^{-1} ,
\eql{pi-decay-current-current}
\eeq
where the estimate \eq{f-pi} for $f_\hpi$ has been used, and we have
assumed $m_{f_i}, m_{f'_j} \ll m_\hpi$. Note that the necessity for the
chirality flip gives the suppression $m_{f_i}^2 + m_{f'_j}^2 /
m_{\hpi}^2$, exactly analogous to the dominance of $\pi^+ \to \mu^+ +
\nu_\mu$ over $\pi^+ \to e^+ + \nu_e$ in the QED-QCD system. On the
other hand, the rate for $\hpi_{\bar{I}J} \to f_i f'_j$ is given (again
up to a group theory factor) by
\beq
  \Ga_{\hpi \to f_i f'_j}
  &=& \fr{C_{\hpi f_i f'_j}^2}{16\pi} \fr{f_\hpi^2 m_\hrho^2}{M^4} \, m_\hpi  \nn\\
  &\approx& \lt[ 100\>\m \times
            \fr{1}{C_{\hpi f_i f'_j}^2 N}
            \fr{100\,\Gev}{m_\hpi}
            \lt( \fr{\Tev}{m_\hrho} \rt)^{\!4} \!
            \lt( \fr{M}{10^4\,\Tev} \rt)^{\!4} \rt]^{-1} ,
\eql{pi-decay-scalar-scalar}
\eeq
where again we have assumed $m_{f_i}, m_{f'_j} \ll m_\hpi$.

To summarize, we can safely conclude that the $\hpi$ carrying net species number are stable on collider time scales. It is possible for such $\hpi$ to
decay within the detector to SM fermion pairs, but in that case avoiding constraints from flavor violation requires that a nontrivial flavor structure be
imposed on the coefficients $C_{\hpi \bar{f}_i f'_j}$ and $C_{\hpi f_i f'_j}$. Since the decay rates depend on the scale $M$ with a high power, displaced
vertices would be visible in a narrow range for $M$ only.

\subsection{Typical Constraints}

Let us now discuss a few generic constraints on the Vectorlike
Confinement scenario. First, integrating out the $\hrho$ which mix with
SM gauge bosons yields {\it flavor-universal} 4-fermion operators, which
are constrained by ``fermion compositeness'' searches. From the
couplings \eq{rho-f-f}, the coefficients of the 4-fermion operators
become (up to $\cO(1)$ SM group theory factors)
\beq
  \fr{\vep_G^2 g_G^2}{m_\hrho^2}
  \approx \fr{N \al^{2}_{G}}{m_\hrho^2}
  \equiv  \fr{2\pi}{\La_{\rm comp}^2}  \,,
\eeq
where we have used \eq{mixing}. $\La_{\rm comp}$ can then be directly
compared with the bounds listed in PDG \cite{PDG}. The most severely
constrained operators are of the form $eeqq$, which arise if there is a
$\hrho$ that mixes with the $Z$, and for $N=3$ the bounds require us to
take $m_\hrho > 500$ GeV.

When a model contains $\hpi$'s that decay to $W$'s, $Z$'s or $\gamma$'s,
it can be constrained by the searches at the Tevatron for such states
\cite{:2008wg}. It turns out that for the masses above 250-300 GeV the
Tevatron does not have sensitivity to discover such $\hpi$ because they
are only electroweak charged and must be pair-produced, requiring higher
center-of-mass energies. Note that since these decays are prompt, the
Tevatron searches for long-lived particles decaying to $\gamma$'s
\cite{Abazov:2008zm} or $Z$'s \cite{Scott:2004wz} do not apply.

For long-lived $\hpi$ that are charged or colored, we will always take
the scale of the operators in \eq{L-pi-f-f} such that the lifetimes are
small compared to cosmological timescales. This ensures that there is no
conflict with the strong constraints from ocean-bottom searches
\cite{Smith:1979rz,Hemmick:1989ns,Verkerk:1991jf,Yamagata:1993jq,Smith:1982qu}.
Collider constraints for the production of collider-stable charged or
colored particles will be mentioned in the discussion of representative
models in the next section, the result being that the mass scales in our
models lie above the exclusion limits.

The $\hrho$ which couple to SM fermions via \eq{rho-f-f} appear as
resonances in dijet and/or dilepton channels. However, the $\hrho$ decay
to $\hpi\hpi$ \eq{rho-width} completely dominates over the decays to SM
fermion pairs \eq{rho-width-to-ff}. The branching fraction to SM
fermions is largest for the $\hrho$ which mixes with a gluon, which
therefore can decay to quarks. Even in that case, however, the Tevatron
dijet resonance searches \cite{Hatakeyama:2008tz} cannot exclude such a
$\hrho$, as shown in \cite{coloron}. As for dileptons, the fact that it
has to go through a $\hrho$ that mixes with a $Z$ or $\gamma$ reduces
the cross section, and together with the small branching fraction, this
is easily consistent with Tevatron resonance searches
\cite{CDF-dielectron,Abazov:2008zm,Aaltonen:2008ah,D0-dielectron,D0-dimuon}.

\subsection{Summary}
\label{sec:shortcut}

\begin{figure}
\begin{center}
\includegraphics[width=0.9\linewidth]{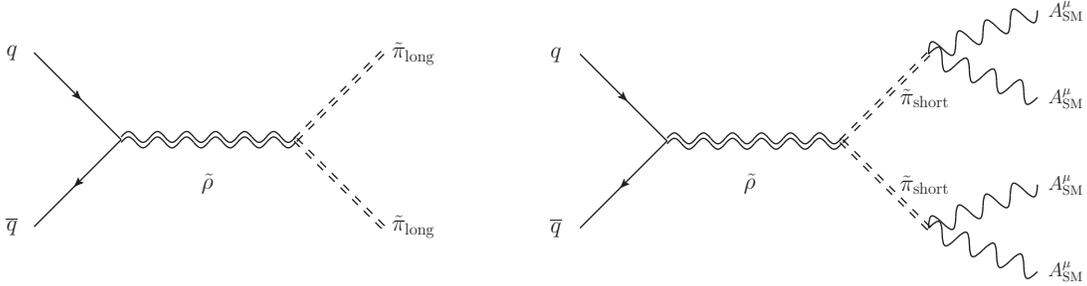}
\end{center}
\caption{The resonant production of hyper-$\rho$'s and their subsequent
decays to a pair of hyperpions. If the hyperpions are of the short-lived
type they will decay promptly to a pair of SM gauge bosons whereas if
they are of the long-lived type they will register as long-lived massive
charged/colored particles.}
\label{fig:production}
\end{figure}

Before we go on to illustrate some of the generic phenomenological features of Vectorlike Confinement, we want to summarize the types of processes
generated by the effective Lagrangian in \eq{pheno-lag}. At the LHC, hyper-$\rho$'s can be resonantly produced from a $q$-$\overline{q}$ initial state and
decay to a pair of hyperpions, which is illustrated in figure \ref{fig:production}. The $\hrho$ that can be resonantly produced can be thought of as
massive copies of the SM gauge bosons (photons, $W$'s, $Z$'s or gluons). Note that the hyperpions that are charged under the SM will also be pair produced
through Drell-Yan processes, in fact this is in general the dominant production mechanism. Once produced, the short-lived hyperpions will decay promptly
to a pair of SM gauge bosons while the long-lived hyperpions will generically pass through the detector without decaying, in which case they will appear
as CHAMP's and R-hadrons, or may be invisible in some cases. It is however possible for the long-lived pions to decay to SM fermion pairs within the
detector if the terms in \eq{L-pi-f-f} leading to their decay incorporate a nontrivial flavor structure.

%%%%%%%%%%%%%%%%%%%%%%%%%%%%%%%%
%%%%%%%%%%%%%%%%%%%%%%%%%%%%%%%%
\section{Phenomenological Highlights}
\label{sec:pheno}

In this section, we would like to go through various phenomenological
features that generically arise in Vectorlike Confinement. In
particular, the existence of short-lived and long-lived
pseudo-Goldstones leads to multiple gauge bosons and collider stable
charged particles in the final state, respectively. Depending on whether
the long-lived particles carry SU(3) color charge or not, they are
classified as R-hadrons or CHAMPs, respectively. They are always
produced in pairs, and when they are created by the decay of a $\hrho$,
it becomes possible to reconstruct the primary resonance, which is a
novel signature.

It is possible for the long-lived $\hpi$ to still decay within collider
timescales if the nonrenormalizable interactions that allow them to
decay have a relatively low scale, which can lead to narrow resonances
of various SM fermion pairs such as dileptons, diquarks or leptoquarks.
(Recall that this is only possible with an extra flavor structure in the
coefficients appearing in \eq{L-pi-f-f}.) Finally, if one (or more) of
the SM singlet $\hpi$ is extremely long-lived or exactly stable, it is a
(possibly decaying) dark matter candidate.

For each signature type we will choose a simple model of Vectorlike
Confinement in which they occur, but more intricate setups are certainly
possible. For each model we discuss, we will also briefly comment on how
hyperbaryons can decay. Hyperbaryon phenomenology depends strongly on
the number of hypercolors and we will not attempt to cover all
possibilities in complete generality. Instead, for each model we
discuss, we will explicitly write down an operator for a specific choice
of the hypercolor group that violates hyperbaryon number so its
inclusion in the Lagrangian can lead to hyperbaryon decay.

%%%%%%%%%%%%%%%%%%%%%%%%%%%%%%%%
%%%%%%%%%%%%%%%%%%%%%%%%%%%%%%%%
\subsection{Multiple Gauge-boson Production}

Except for the case of all $\Psi$ being SM singlets, Vectorlike
Confinement phenomenology always includes one (or more) $\hrho$ that
mixes with a SM gauge boson as well as a $\hpi$ in the same SM
representation. Such a $\hpi$ decays promptly to a pair of SM gauge
bosons via \eq{pi-F-F}. Therefore, for {\it any} SM charge assignments
of the $\Psi$, when such a $\hrho$ is resonantly produced from a
$q$-$\bar{q}$ initial state via \eq{rho-f-f}, one of its decay modes is
always to a pair of $\hpi$ which subsequently decay to four SM gauge
bosons. This final state is robust in all models of Vectorlike
Confinement.

Such a signature with four gluons was studied in \cite{coloron} for the
3-flavor, 1-species model with $\Psi \sim ({\bf 3}, {\bf 3})$ under
$\SU(3)_\text{HC} \otimes \SU(3)_\text{C}$. In this case there is no
coupling to leptons since the hyper-sector couple to color only, and the
$\hrho$, which mixes with a gluon, can be as light as 350 GeV, and has
been shown to be discoverable in existing Tevatron data.

Naively, one might think that an analogous model for electroweak
interaction, i.e.\ the 2-flavor 1-species model with $\Psi \sim ({\bf
N}, {\bf 2})$ under $\SU(N)_\text{HC} \otimes \SU(2)_\text{W}$, should
give signatures with four electroweak gauge bosons. However, this is
incorrect for a subtle reason specific to this model, which we discuss
in Appendix \ref{app:2-flavor-model}. The signature of multiple
electroweak gauge boson production is retained if one adds one more
hyperflavor to the model, which is the case we will discuss next. Since
this addition also causes the existence of a di-CHAMP resonance, we will
investigate both types of signatures in the next subsection
\Sec{di-Champ}.

%%%%%%%%%%%%%%%%%%%%%%%%%%%%%%%%
%%%%%%%%%%%%%%%%%%%%%%%%%%%%%%%%
\subsection{A Model of multi EW-boson Production and Di-CHAMP Resonances}
\Secl{di-Champ}

In order to have a long-lived $\hpi$ in the spectrum, the hyperflavors
need to arrange themselves into at least two species. For this purpose
we will choose to work with the following model with 3 hyperflavors
which arrange themselves into 2-species, a weak doublet and a singlet
with hypercharge:
\beq
\begin{tabular}{c|c||c|c|c}
  & $\SU(N)_{\rm HC}$ & $\SU(3)_{\rm C}$ & $\SU(2)_{\rm W}$ & $\U(1)_{\rm Y}$  \\
\hline
$\Psi_{\bf 2}$ & $\Box$ & ${\bf 1}$ & $\Box$    & -1/2  \\
$\Psi_{\bf 1}$ & $\Box$ & ${\bf 1}$ & ${\bf 1}$ & 1
\end{tabular}
\eql{2x1in3}
\eeq
Note that $\Psi_{\bf 1,2}$ are Dirac fermions.

\subsubsection{Spectrum}

Let us begin by listing the states in the spectrum. In order to display
the quantum numbers of each state we will use the notation
$(r_{\SU(3)_{\rm C}},r_{\SU(2)_{\rm W}})_{\U(1)_{\rm Y}}$. The $\hrho$
which mix with SM gauge bosons are
\beq
  (W^{\prime 0}, W^{\prime\pm}) &\sim& ({\bf 1}, {\bf 3})_0  \,,\nn\\
  B'                      &\sim& ({\bf 1}, {\bf 1})_0  \,,
\eeq
while the rest of $\hrho$ transform as
\beq
  (\hrho^{++}, \hrho^{+}) &\sim& ({\bf 1}, {\bf 2})_{3/2}  \,,\nn\\
  \hrho''_0               &\sim& ({\bf 1}, {\bf 1})_0  \,,
\eeq
where the $\hrho''_0$ corresponds to the hyperbaryon $\U(1)_\text{HB}$
current. The $\hpi$'s fill in the following multiplets:
\beq
  (\hpi^0, \hpi^\pm)    &\sim& ({\bf 1}, {\bf 3})_0  \,,\nn\\
  (\hK^{++}, \hK^{+})   &\sim& ({\bf 1}, {\bf 2})_{3/2}  \,,\nn\\
  \hpi'_0               &\sim& ({\bf 1}, {\bf 1})_0  \,.
\eeq
The $\hpi$ masses from SM gauge loops \eq{pi-mass-gauge} before EWSB are
\beq
  m_{\hpi^0, \hpi^\pm}  &\simeq& 0.13 c \, m_\rho  \,,\nn\\
  m_{\hK^{++}, \hK^{+}} &\simeq& 0.11 c \, m_\rho  \,,
\eql{pi-spec-2x1in3}
\eeq
where the coefficient $c$ was defined in equation
(\ref{eq:pi-mass-gauge}) and has the value 1.1 for $N=3=F$ as we know
from QCD. The SM-singlet $\hpi'_0$ does not get a mass from SM gauge
loops, so $m_{\hpi'_0}$ arises entirely from hyperquark masses
\eq{pi-mass-from-mhq}. After EWSB, the $\hpi^\pm$-$\hpi^0$ and
$\hK^{++}$-$\hK^+$ mass splittings are given by \eq{mass-split-IR} as
\beq
  m_{\hpi^\pm} - m_{\hpi^0} &\approx& 170\>\Mev  \,,\eql{166MeV}  \\
  m_{\hK^{++}} - m_{\hK^+}  &\approx& 1.1\>\Gev  \,.\eql{1.1GeV}
\eeq

\subsubsection{Model specific constraints}
\label{sec:constraints for twoplusone model}

\begin{figure}[t]
\begin{center}
\includegraphics[width=0.5\linewidth]{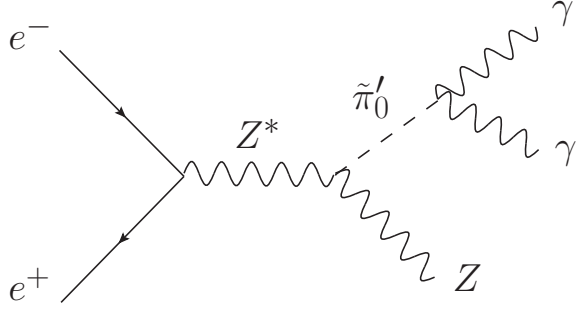}
\end{center}
\caption{Emission of an $\hpi^{0}$ in a Drell-Yan process and subsequent
decay to a pair of photons is found to be consistent with low mass Higgs
boson searches at LEP.}
\label{fig:ee2Za}
\end{figure}

The SM-singlet $\hpi'_0$ shares many properties with axions, such as the coupling to two photons via $\hpi'_0 F \td{F}$ from \eq{pi-F-F}. Unlike the
ordinary Peccei-Quinn axion, however, its couplings to SM fermions are induced only at two-loop level or via nonrenormalizable operators that couple
hyperquarks to SM fermions. (It still couples to nucleons with a similar strength as the ordinary axions because it mixes with a $\pi^0$). The two-loop
suppression, however, is not enough to evade astrophysical bounds such as supernova cooling if $m_{\hpi'_0} \lsim 100$ MeV. On the other hand, it is
enough to account for the absence ($\lsim 10^{-10}$ \cite{PDG}) of the decays $K^+ \to \pi^+ A^0 \to \pi^+ e^+e^-$ or $\to \pi^+ \mu^+ \mu^-$, where $A^0$
is an axion, because the two-loop suppression factor enters {\it twice} in the amplitudes of these processes. Therefore, we take $m_{\hpi'_0}$ to be at
least $\sim \cO(100)$ MeV, corresponding to hyperquark masses of $\gsim \cO(10)$ keV for $m_\hrho \sim$ TeV. Then the constraints from beam-dump
experiments which looked for axion-like particles are also evaded \cite{beam-dump}\footnote{A recent interest in studying ``dark'' $\U(1)$ gauge bosons
has also led to a reanalysis of these beam-dump experiments \cite{Bjorken:2009mm} in terms of parameters that are better suited for a light vector
particle rather than an axion. Future experiments proposed in \cite{Bjorken:2009mm} will almost certainly improve the bounds on $\hpi'_0$ as well.}.

If $m_{\hpi'_0} < 10$ GeV, a $\Upsilon$ can decay into a $\ga$ and a
$\hpi'_0$ which then decays promptly to two $\ga$'s. This could in
principle be observable at BaBar \cite{BaBar-Upsilon}, but since the
$\hpi'_0$-$\ga$-$\ga$ vertex \eq{pi-F-F} is one-loop suppressed, it only
amounts to less than one event. Similarly, since the $\hpi'_0$-$Z$-$Z$
coupling is also one-loop suppressed, the production of a $\hpi'_0$ in
association with a $Z$ at lepton colliders (see \fig{ee2Za}) is highly
suppressed, and therefore falls way below the bounds from the LEP Higgs
searches in the similar channel \cite{Rosca:2002me}.

\subsubsection{Decays and collider phenomenology}

Let us now discuss the phenomenology of the $\hrho$ and non-SM-singlet
$\hpi$. In colliders, a $W'$ can be resonantly produced from a
$q$-$\bar{q}$ initial state, and will dominantly decay via
\eq{rho-pi-pi} to
\beq
  W^{\prime 0}  &\stackrel{2/3}{\too}& \hpi^{+} + \hpi^{-}   \nn\\
             &\stackrel{1/6}{\too}& \hK^{++} + \hK^{--}   \nn\\
             &\stackrel{1/6}{\too}& \hK^{+} + \hK^{-}     \,,\nn\\
  W^{\prime +}  &\stackrel{2/3}{\too}& \hpi^{0} + \hpi^{+}   \nn\\
             &\stackrel{1/3}{\too}& \hK^{++} + \hK^{-}    \,.
\eeq
Similarly, a $B'$ can also be resonantly produced and dominantly decays
via \eq{rho-pi-pi} as
\beq
  B' &\stackrel{1/2}{\too}& \hK^{++} + \hK^{--}  \nn\\
     &\stackrel{1/2}{\too}& \hK^{+} + \hK^{-}    \,.
\eeq
(The branching fractions above are approximate as the differences of the
hyperpion masses are ignored.)

The hyperpions can be produced from Drell-Yan processes or through the
decays above. The $\hpi^\pm$ and $\hpi^0$ decay promptly to electroweak
gauge bosons via the vertex \eq{pi-F-F}:
\beq
  \hpi^\pm &\too& W^\pm Z \,,\> W^\pm \ga            \,,\nn\\
  \hpi^0   &\too& 2Z      \,,\> Z\ga      \,,\> 2\ga \,.
\eql{pi_BFs}
\eeq
(These are actually $W^\pm B$ and $W^3 B$, respectively.) Since the
branching fractions of these decays have only a mild dependence on phase
space, we take a representative value\footnote{This corresponds to
$m_{\hrho}=2.5~\rm{TeV}$ which we will use in \fig{CHAMPres}.} of
$m_{\hpi}=355$ GeV and find that the $\hpi^\pm$ decays to $W\gamma$ 81\%
of the time and to $WZ$ 19\% of the time, while the branching fractions
of the $\hpi^0$ to $\gamma\gamma$,$\gamma Z$ and $ZZ$ are given by 43\%,
29\% and 28\% respectively. We therefore have a very interesting
collider phenomenology with little background. Since these $\hpi$ are
pair-produced where each in turn decays to a pair of SM gauge bosons,
there will be a significant fraction of events with multiple leptons as
well as photons in the final state. In particular, the $\hpi^{0}$
pair-production will give rise to $4\gamma$ events which can be paired
up to reveal the $\hpi^0$ resonances. Similarly, the $\hpi^{\pm}$ pair
production can lead to $\ell^{+}\ell^{-}\gamma\gamma$ + missing
$E_\text{T}$.

In contrast, the $\hK$ carry nonzero species numbers and thus do not
have the couplings \eq{pi-F-F} to gauge bosons. Instead, because of the
mass splitting \eq{1.1GeV}, a $\hK^{++}$ can decay to a $\hK^{+}$, which
is lighter and carries the same species numbers:
\beq
  \hK^{++} &\too& \hK^+ + \pi^+            \,,\quad
                  \hK^+ + K^+              \,,\nn\\
           &&     \hK^+ + e^+   + \nu_e    \,,\quad
                  \hK^+ + \mu^+ + \nu_\mu  \,.
\eql{dim-4-K++decays}
\eeq
These decays are mediated by an off-shell $W$ boson, and the rates are
given by
\beq
 \Ga_{\hK^{++} \to \hK^+ + \pi^+}^{-1}
   &=& 220\>\mum \lt( \fr{1.1\>\Gev}{\De m_\hK} \rt)^{\!3}
       \sqrt{\fr{1 - m_{\pi^+}^2/(1.1\>\Gev)^2}{1 - m_{\pi^+}^2/\De m_\hK^2}}  ,\\
 \Ga_{\hK^{++} \to \hK^+ + K^+}^{-1}
   &=& 4.1\>\mm \lt( \fr{1.1\>\Gev}{\De m_\hK} \rt)^{\!3}
       \sqrt{\fr{1 - m_{K^+}^2/(1.1\>\Gev)^2}{1 - m_{K^+}^2/\De m_\hK^2}}  ,
\eeq
and
\beq
  \Ga_{\hK^{++} \to \hK^+ + e^+ + \nu_e}^{-1}
  &=&      480\>\mum \lt( \fr{1.1\>\Gev}{\De m_\hK} \rt)^{\!5}  \nn\\
  &\simeq& \Ga_{\hK^{++} \to \hK^+ + \mu^+ + \nu_\mu}^{-1}  .
\eeq
Unfortunately, even though these rates are of the right size for
displaced vertices, the outgoing $\pi^\pm$, $e^\pm$, $\mu^\pm$ are all
too soft to be seen in detectors. The bright side of this fact is that
since the energy of the parent $\hK^{\pm\pm}$ is almost entirely carried
away by the daughter $\hK^{\pm}$, a $\hK^{\pm\pm}$ can practically be
identified as a $\hK^\pm$ in collider experiments, effectively
increasing the $\hK^\pm$ production rate.

On the other hand, the $\hK^\pm$ is the lightest particle with nonzero
species numbers, and therefore can only decay via a non-renormalizable
operator that violate species numbers, as discussed in \Sec{accidental}.
The leading such operator is given by the current-current operator
\eq{current-current} with $(\Psi_J, \Psi_I, f_i, f_j) = (\Psi_{\bf 1},
\Psi_{\bf 2}, e^c_i, \ell_j)$. This operator gives rise to
\beq
   \hK^+    &\too& \bar{\ell}_i + \nu_j  \,\\
   \hK^{++} &\too& \bar{\ell}_i + \bar{\ell}_j  \,,
\eeq
with the rates given by \eq{pi-decay-current-current}. Therefore, it is
quite likely that $\hK^\pm$ is collider stable (unless a nontrivial
flavor structure is present in \eq{pi-decay-current-current} to allow
for significantly lower $M$), while the $\hK^{\pm\pm}$ width is
dominated by the weak decays \eq{dim-4-K++decays}.

Thus, we expect a novel collider signature, a ``di-CHAMP resonance,''
where it is possible to reconstruct the parent $W^{\prime 0}$ or $B'$
from a $\hK^+$-$\hK^-$ pair. Here recall that a $\hK^{\pm\pm}$ cannot be
distinguished from a $\hK^{\pm}$, so $W^{\prime 0}$ and $B'$ decays to
$\hK^{\pm\pm}$ as well as $W'^\pm$ production also contribute to the
signal. Since CHAMPs have no irreducible backgrounds, we do not need a
large luminosity for CHAMP analysis. Note, however, that the CHAMPs can
also be produced through Drell-Yan processes, which tends to be larger
than the production through a $\hrho$ resonance, due to the large mass
of the $\hrho$. In \fig{CHAMPres}, we plot the CHAMP production cross
section as a function of the pair invariant mass at the LHC
($\sqrt{s}=14$ TeV), with the number of hypercolors $N=3$ (thus taking
$c=1.1$ in \eq{pi-spec-2x1in3}) for $m_\hrho = 2.5$ TeV
($m_\hK=3.0\times10^{2}$ GeV) and $m_\hrho = 4.0$ TeV
($m_\hK=4.8\times10^{2}$ GeV). For this plot, the parameter values
$g_{\hrho\hpi\hpi}=6$, $\vep_{W'}=0.1$ and $\vep_{B'}=0.1$ were used.
(Note that the latter includes a factor of $\sqrt{3}$ due to the
normalization of SM hypercharge relative to the corresponding
hyperflavor generator $T^{8}$.) One sees that even though most events
are induced by Drell-Yan processes, there is enough statistics in the
lighter case for the parent $\hrho$ resonance to be clearly noticeable,
given that there is no irreducible background. Furthermore, it is
possible to analyze the angular distribution of the CHAMP pairs and see
that the primary resonance is indeed a spin-1 particle decaying to a
pair of scalars. In particular, one can do this by keeping only the high
invariant mass events, subtracting the Drell-Yan ``background'', which
has the same angular distribution.

\begin{figure}[t]
\includegraphics[width=0.95\linewidth]{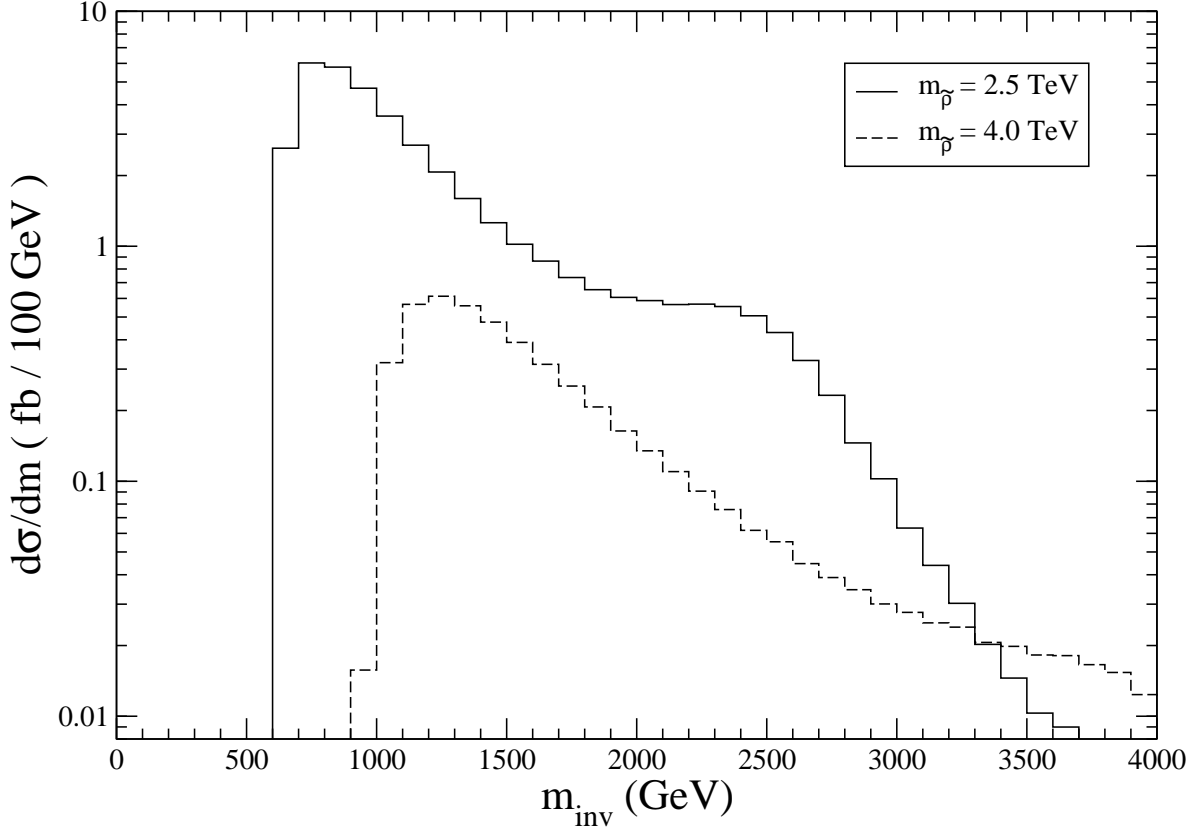}
\caption{The differential cross section for the pair production of CHAMPs for $\sqrt{s}=14$ TeV, $m_{\hrho}=2.5~\rm{TeV}$ and $m_{\hrho}=4.0~\rm{TeV}$ at
the LHC. While most of the CHAMPs come from Drell-Yan processes, at high enough energies resonant $\hrho$ production becomes the dominant source.}
\figl{CHAMPres}
\end{figure}

In principle, there is also a possibility of pair-producing $\hrho$'s.
Note, however, that the large mass gap \eq{pi-spec-2x1in3} between the
$\hpi$ and $\hrho$ implies $m_\hrho \sim$ a few TeV. Having only
electroweak gauge interactions, a pair of $\hrho$'s in this mass range
is challenging at the LHC.

Finally, all CHAMP and R-hadron searches at the Tevatron
\cite{CHAMPs-Rhadrons, Aaltonen:2009kea} limit the production
cross-section of the $\hpi$ to be less than about $0.1$ pb. As in the
LHC case above, they are dominantly produced by SM gauge interactions,
and the $0.1$ pb bound on the cross section translates to a lower bound
on the CHAMP mass of about 250-300 GeV.

\subsubsection{Hyperbaryons}

Before moving on to the next model, we would like to comment briefly on
hyperbaryon decays in this model. In the specific case where $N_{\rm
HC}=4$, one can include in the Lagrangian the following operator (among
others) that violates hyperbaryon number:
\begin{equation}
{\mathcal O}_{\rm HB}=\epsilon^{ABCD}\left(\Psi^{\rm T}_{{\bf 2}A}\,{\mathcal C}\,\Psi_{{\bf 2}B}\right)\left(h\,\Psi^{\rm T}_{{\bf 2}C}\,{\mathcal
C}\,\Psi_{{\bf 1}D}\right)+{\rm h.c.}\label{eq:HBviolating2}
\end{equation}
where the uppercase letters denote hypercolor indices and ${\mathcal C}$
is the charge conjugation matrix. The $\SU(2)_{\rm W}$ indices are
implicitly contracted within each bracket, between a pair of $\Psi_{\bf
2}$'s or a $\Psi_{\bf 2}$ and a SM Higgs. One can check that the
hypercharges add to zero. ${\mathcal O}_{\rm HB}$ can lead to the decay
of a hyperbaryon to a number of hyperpions and a component of the Higgs
doublet.

For different choices of $N_{\rm HC}$ one can write down similar
operators by replacing some of the $\Psi$'s by a judicious choice of SM
fermions and adding Higgs insertions such that all SM gauge indices can
be contracted. Note that the mass dimension of the operator will in
general depend on $N_{\rm HC}$.

%%%%%%%%%%%%%%%%%%%%%%%%%%%%%%%%
%%%%%%%%%%%%%%%%%%%%%%%%%%%%%%%%
\subsection{A Model of Di-R-hadron Resonances}
\Secl{di-R-hadron}

The appearance of R-hadrons in Vectorlike Confinement theories is quite
generic, a necessary and sufficient condition being just the existence
of two (or more) species, at least one of which carries QCD color. In
order to illustrate the relevant phenomenology we choose the following
minimal model with four flavors that arrange themselves into two
species, one color triplet and a singlet with hypercharge:
\beq
\begin{tabular}{c|c||c|c|c}
  & $\SU(N)_{\rm HC}$ & $\SU(3)_{\rm C}$ & $\SU(2)_{\rm W}$ & $\U(1)_{\rm Y}$  \\
\hline
$\Psi_{\bf 3}$ & $\Box$ & $\Box$    & ${\bf 1}$ & -1/3  \\
$\Psi_{\bf 1}$ & $\Box$ & ${\bf 1}$ & ${\bf 1}$ & 1
\end{tabular}
\eql{3x1in4}
\eeq
Note that $\Psi_{\bf 1,3}$ are Dirac fermions.

\subsubsection{Spectrum}

The $\hrho$ and $\hpi$ spectra are as follows. The $\hrho$ which mix
with SM gauge bosons are
\beq
  g' &\sim& ({\bf 8}, {\bf 1})_0  \,,\nn\\
  B' &\sim& ({\bf 1}, {\bf 1})_0  \,,
\eeq
while the rest of $\hrho$'s transform as
\beq
  \hrho_{\bf 3} &\sim& ({\bf 3}, {\bf 1})_{-4/3}  \,,\nn\\
  \hrho''_0      &\sim& ({\bf 1}, {\bf 1})_0      \,,
\eeq
where the $\hrho''_0$ corresponds to the hyperbaryon $\U(1)_\text{HB}$
current. The $\hpi$ fill in the following multiplets.
\beq
  \hpi_{\bf 8} &\sim& ({\bf 8}, {\bf 1})_0      \,,\nn\\
  \hpi_{\bf 3} &\sim& ({\bf 3}, {\bf 1})_{-4/3}  \,,\nn\\
  \hpi'_0      &\sim& ({\bf 1}, {\bf 1})_0      \,.
\eeq
The $\hpi$ masses from SM gauge loops \eq{pi-mass-gauge} are
\beq
  m_{\hpi_{\bf 8}} &\simeq& 0.29 c \, m_\hrho  \,,\nn\\
  m_{\hpi_{\bf 3}} &\simeq& 0.20 c \, m_\hrho  \,,
\eql{pi-mass-3x1in4}
\eeq
while the singlet $\hpi'_0$ does not get a mass from SM gauge loops, so
$m_{\hpi'_0}$ arises entirely from hyperquark masses
\eq{pi-mass-from-mhq}.

\subsubsection{Model specific constraints}

As we discussed in section \ref{sec:constraints for twoplusone model},
we take $m_{\hpi'_0} \gsim \cO(100)$ MeV. But unlike the $\hpi'_0$
there, the $\hpi'_0$ here not only couples to two photons but also to
two gluons via \eq{pi-F-F}. Thus, a $\hpi'_0$ can be resonantly produced
from $gg$ at hadron colliders, which subsequently decays to $gg$
($\approx 99\%$) or $\ga$-$\ga$ ($\approx 1\%$). Just like the $gg \to
\hpi_{\bf 8} \to gg$ case discussed in \cite{coloron}, the $gg \to
\hpi'_0 \to gg$ dijet process here is completely overwhelmed by QCD
dijet processes, due to the one-loop suppression in the vertex
\eq{pi-F-F}. On the other hand, the $gg \to \hpi'_0 \to \ga\ga$ could be
clean enough, despite its low branching fraction. At the Tevatron,
inclusive di-photon searches have been performed \cite{Acosta:2004sn},
but it turns out that the $\hpi'_0$ is not ruled out for $m_{\hpi'_0} <
100$ GeV. (We will not consider $m_{\hpi'_0} \gsim 100$ GeV, as this
cannot be obtained from hyperquark masses $\ll m_\hrho$.) See Appendix
\ref{app:axion-collider-const} for the detail of the analysis. Finally,
at Babar, a $\Upsilon$ can decay into a photon and a $\hpi'_0$ which
subsequently decays promptly to two gluons. While this dominates over
the two-photon mode discussed in \Sec{di-Champ}, it would be completely
buried under SM backgrounds.

\subsubsection{Decays and collider phenomenology}

Let us move on to the phenomenology of the $\hrho$ and the non-singlet
$\hpi$. The $g'$ and $B'$ can be produced resonantly at colliders and
decay via \eq{rho-pi-pi}:
\beq
  g' &\stackrel{3/4}{\too}& \hpi_{\bf 8} + \hpi_{\bf 8}        \nn\\
     &\stackrel{1/4}{\too}& \hpi_{\bf 3} + \wba{\hpi_{\bf 3}}  \,,\nn\\
  B' &\too&                 \hpi_{\bf 3} + \wba{\hpi_{\bf 3}}  \,,
\eql{gprimeBFs}
\eeq
(These branching fractions are approximate as the $\hpi$ mass
differences are ignored.)

The $\hpi_{\bf 8}$ decays promptly via \eq{pi-F-F} as
\beq
  \hpi_{\bf 8} &\too& 2g \,,\> gZ   \,,\> g\ga  \,.
\eeq
\begin{figure}[t]
\includegraphics[width=0.95\linewidth]{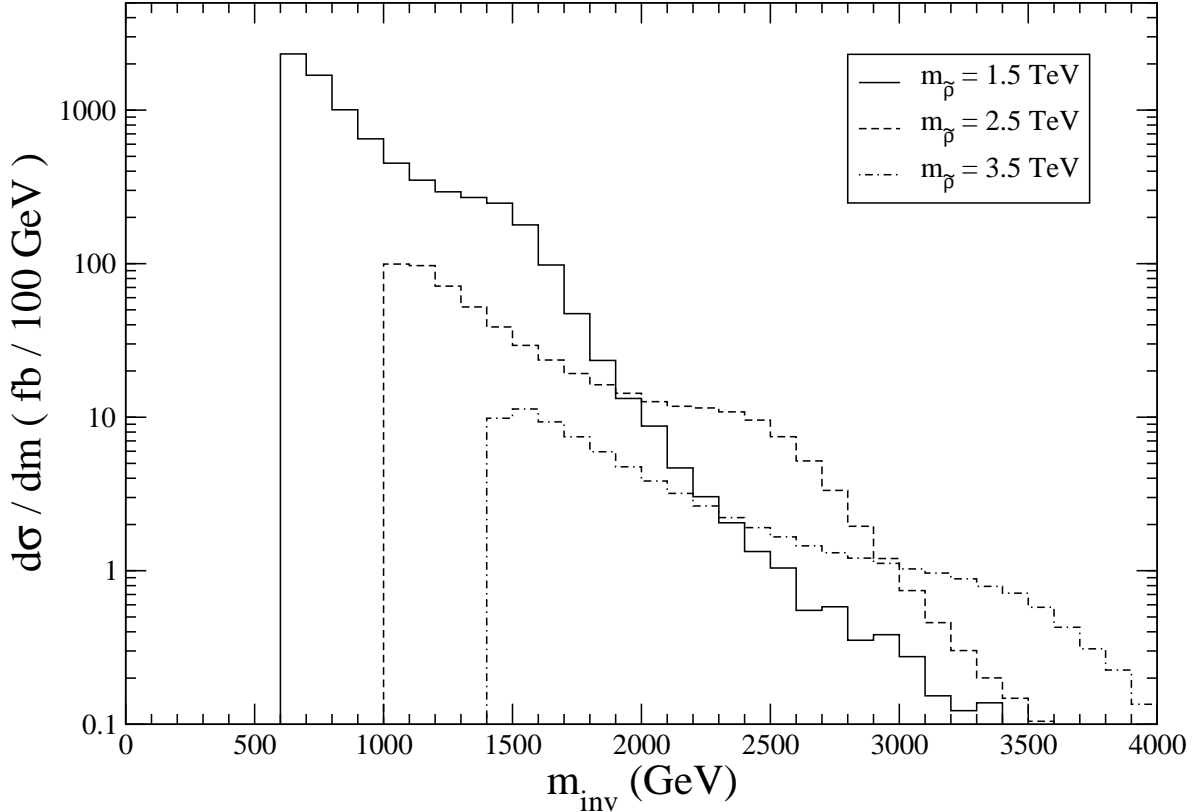}
\caption{The differential cross section for the pair production of
R-hadrons at the LHC ($\sqrt{s}=14$) TeV for three choices of $m_{\hrho}$
corresponding to an $m_{\hpi_{\bf 3}}$ of 300,500 and 700~GeV. At high energies,
decays of resonantly produced $\hrho$ dominate the production.}\label{fig:Rhadron}
\end{figure}
The $\hpi_{\bf 3}$ requires a nonrenormalizable operator to decay, as it carries nonzero species number. The leading operator for $\hpi_{\bf 3}$ decay is
given by the ``scalar-scalar'' operator \eq{scalar-scalar} with $(\Psi_J, \Psi_I, f_i, f_j) = (\Psi_{\bf 1}, \Psi_{\bf 3}, e^c_i, d^c_j)$. This induces
the decay
\beq
  \hpi_{\bf 3} \too \ell_i + d_j
\eeq
with the rate given by \eq{pi-decay-scalar-scalar}. Therefore, we expect that the $\hpi_{\bf 3}$ is collider stable (unless an extra flavor symmetry is
introduced to significantly lower $M$). Once produced, it will hadronize rapidly into an R-hadron, sometimes neutral and sometimes charged, in roughly
equal numbers. If both R-hadrons from a $g'$ or $B'$ decay end up charged, then we again have a resonance in a pair of long lived charged particles.
Although the $\hrho$ width in this case is slightly larger than the case of \Sec{di-Champ} due to larger $F$ in \eq{rho-width}), the resonant $\hrho$ can
now be lighter, since $m_{\hpi}/m_{\hrho}$ from \eq{pi-mass-gauge} is larger. Therefore, we expect a larger cross section for the resonant production of
$\hpi_{\bf 3}$'s than that of $\hK^\pm$'s in \Sec{di-Champ}, which is illustrated in figure \ref{fig:Rhadron}. In this plot the parameter values
$g_{\hrho\hpi\hpi}=6$, $c=1$ and $\vep_{g}=0.2$ were used. Again, it should be interesting to analyze the angular distribution of the R-hadron pairs and
observe that the primary resonance is indeed spin-1.

Note that for a given $\m_{\hpi}$, $m_{\hrho}$ can be lighter compared
to the previous model, so that pair-producing $g'$'s and $\hrho_{\bf
3}$'s might be interesting as well. Such pair-production of $g'$ was
shown to be discoverable in the multijet channel \cite{coloron_LHC}.
Including the R-hadrons in the decay products can make the signal quite
spectacular, with the production of 4 R-hadrons with an $\cO(\fb)$
cross-section.

\subsubsection{Hyperbaryons}

Similar to the model presented previously, we can write down a Lorentz
invariant, gauge singlet operator that allows the hyperbaryons to decay
in this model. Once again, we content ourselves with a specific example
for the choice of the hypercolor group, in particular for $N=4$ one can
write down
\begin{equation}
{\mathcal O}_{\rm HB}=\epsilon^{ABCD}\left(\Psi^{\rm T}_{{\bf 3}A}\,{\mathcal C}\,\Psi_{{\bf 3}B}\right)\left(\Psi^{\rm T}_{{\bf 3}C}\,{\mathcal
C}\,\Psi_{{\bf 1}D}\right)+{\rm h.c.}\label{eq:HBviolating3}
\end{equation}
where the suppressed $\SU(3)_{\rm C}$ indices of the $\Psi_{\bf 3}$ are
contracted with an $\epsilon^{ijk}$. This operator breaks hyperbaryon
number and including it in the Lagrangian makes it possible for
hyperbaryons to decay into a number of $\hpi$. For different choices of
$N_{\rm HC}$, one can trade $\Psi$'s for SM fermions (or add Higgs
insertions) in order to make a gauge singlet.

%%%%%%%%%%%%%%%%%%%%%%%%%%%%%%%%
%%%%%%%%%%%%%%%%%%%%%%%%%%%%%%%%
\subsection{Leptoquarks, Dileptons, and Diquarks (with Displaced Vertices)}

Another collider signature that can arise in a variety of Vectorlike
Confinement models is that of narrow resonances in pairs of SM fermions.
As we discussed in \Sec{4-fermion}, if extra flavor structures are
imposed on the mixed SM-hyper 4-fermion operators \eq{current-current}
and \eq{scalar-scalar}, it is possible to lower the scale $M$ so that
long-lived $\hpi$ decay within the detector, possibly with displaced
vertices.

For example, the $\hK^\pm$ CHAMP of \Sec{di-Champ} can decay inside the
detector to $\tau^+\nu_\tau$. (Note the strong preference of the rate
\eq{pi-decay-current-current} for heavier flavors.) Similarly,
$\hK^{++}$ will promptly decay to $\tau^+\tau^+$, beating the rate
\eq{dim-4-K++decays}. Therefore, the $\hK^{++}$-$\hK^{--}$ pair
production, either from Drell-Yan or from the $W'^0$ decay, will
dominantly end up in 4$\tau$'s. These $\tau$'s are highly boosted, hence
difficult to identify, but it may help to utilize the leptonic $\tau$
decays.

Similarly, if $M$ is sufficiently lowered, the $\hpi_{\bf 3}$ R-hadron of \Sec{di-R-hadron} will decay inside the detector to a charged lepton and a
$d$-type quark, hence behaving as a leptoquark, offering an exciting collider signature. Although the 4-fermion operator responsible for the $\hpi_{\bf
3}$ decay is of the ``scalar-scalar'' form \eq{scalar-scalar}, we again expect that the rate \eq{pi-decay-scalar-scalar} would prefer decays to the 3rd
generation, because flavor violation constraints requires that this operator should be small for the 1st and 2nd generations. The Tevatron has set the
lower limits on the mass of such ``3rd generation leptoquarks'' at 250-300 GeV \cite{Direct-LQ}.

%%%%%%%%%%%%%%%%%%%%%%%%%%%%%%%%
%%%%%%%%%%%%%%%%%%%%%%%%%%%%%%%%
\subsection{Dark Matter, SUSY Look-alikes and Consistency with Unification}

Vectorlike Confinement models can readily contain dark matter candidates. There are two generic reasons for the existence of extremely long-lived or
stable $\hpi$. The first possibility is that one (or more) species numbers remain unbroken by nonrenormalizable interactions. A simple example is given by
the following model with two species:
\beq
\begin{tabular}{c|c||c|c|c}
  & $\SU(N)_{\rm HC}$ & $\SU(3)_{\rm C}$ & $\SU(2)_{\rm W}$ & $\U(1)_{\rm Y}$  \\
\hline
$\Psi_{\bf 1}$ & $\Box$ & ${\bf 1}$ & ${\bf 1}$ & 1  \\
$\Psi_{\bf 3}$ & $\Box$ & ${\bf 1}$ & ${\bf 3}$ & 1
\end{tabular}
\eeq
This model contains, in addition to several $\hpi$ of the short-lived type, a long-lived {\em complex} $\SU(2)$ triplet hyperpion, with charged and
neutral components split by $170~{\rm MeV}$ after electroweak symmetry breaking. This complex scalar triplet made out of a $\Psi_{\bf 1}$ and a $\Psi_{\bf
3}$ (which we will denote as $\hpi_{13}$) can be stabilized by both an unbroken $\Psi_{\bf 1}$ or an unbroken $\Psi_{\bf 3}$ number\footnote{In an earlier version of this work, the hypercharge of $\Psi_{\bf 1,3}$ was assigned to be zero, leading to additional stable bound states due to a mechanism similar to the one described in appendix \ref{app:2-flavor-model}. We thank Richard Hill for pointing this out to us.}. More precisely, the
neutral (but complex) component $\hpi_{13}^{0}$ is exactly stable, while the two copies of $\hpi_{13}^{\pm}$ can decay to $\hpi_{13}^{0}$ via the SM weak
interactions. The $\hpi_{13}^{0}$ is therefore a WIMP and a dark matter candidate. When the mass of this WIMP is $\sim 2.0~\rm{TeV}$, it has the correct
relic density \cite{minimalDM}. Since the lightest hyperbaryon in this model (made out of $N_{\rm HC}$ copies of $\Psi_{\bf 1}$) is a SM singlet, it is
straightforward to  write terms that make it decay, for instance to a pair of Higgs bosons when $N_{\rm HC}$ is even.

The second possibility arises if the lightest long-lived $\hpi$ has quantum numbers such that no operator of the form (\ref{eq:pi_decay_op_scalar}) or
(\ref{eq:pi_decay_op_vector}) can be written down. While such $\hpi$ may still decay, this would have to occur through operators with high mass dimension,
and lifetimes much larger than the age of the universe become plausible. Such models may therefore contain candidates for decaying dark matter.

One specific possibility we wish to mention is that of a ``spectator'' species, where one species $\Psi_{\rm spec}$ lives in an exotic SM representation
(e.g. a {\bf 4} of $\SU(2)_{\rm W}$). In this case it is possible that various $\hpi$ containing $\Psi_{\rm spec}$ decay into one another, giving rise to
(slow) cascades, until the lightest state containing $\Psi_{\rm spec}$ is reached, which, if neutral, is a dark matter candidate.

To highlight one more interesting phenomenological feature that can arise in Vectorlike Confinement models, consider the following model with 3-species
(6-flavors):
\beq
\begin{tabular}{c|c||c|c|c}
  & $\SU(N)_{\rm HC}$ & $\SU(3)_{\rm C}$ & $\SU(2)_{\rm W}$ & $\U(1)_{\rm Y}$  \\
\hline
$\Psi_{\bf 3}$ & $\Box$ & $\Box$    & ${\bf 1}$ & -1/3  \\
$\Psi_{\bf 2}$ & $\Box$ & ${\bf 1}$ & $\Box$    & 1/2  \\
$\Psi_{\bf 1}$ & $\Box$ & ${\bf 1}$ & ${\bf 1}$ & 0
\end{tabular}
\eeq
Note that $\Psi_{\bf 1,2,3}$ are Dirac fermions\footnote{Contrary to our generic assumptions, this particular setup allows the inclusion in the Lagrangian
of a Yukawa coupling of $\overline{\Psi}_{\bf 2}\Psi_{\bf 1}$ to the Higgs doublet. In order to avoid a conflict with precision electroweak constraints we
will assume here that such a coupling, if it exists, is small, which is just as natural as taking the mass scale appearing in \eq{L-pi-f-f} to be large.}.
One sees that the $\hpi$ corresponding to ``$\wba{\Psi}_{\bf 2} \Psi_{\bf 3}$'', ``$\wba{\Psi}_{\bf 1} \Psi_{\bf 3}$'' and ``$\wba{\Psi}_{\bf 1} \Psi_{\bf
2}$'' are like the left-handed squark doublet, right-handed down squark and left-handed slepton doublet (except that the ``left-handed squark'' has a
wrong hypercharge). Their squared-masses from SM gauge loops \eq{pi-mass-gauge} are $\propto g^2 C_2$, resembling a typical sfermion spectrum in
gauge-mediated SUSY models. Of course, the collider phenomenology of this model differs significantly from that of typical supersymmetric models, because
not only we have a plethora of 4-gauge boson final states but also these ``sfermions'' will be CHAMPs and R-hadrons, unless extra flavor structures are
imposed on the 4-fermion operators \eq{current-current} and \eq{scalar-scalar}. If we do impose extra flavor symmetries and allow the ``sfermions'' to
decay at collider time scales, they decay to a pair of SM fermions. Similar features also appear in the ``SUSY in slow motion'' scenario \cite{SlowSUSY},
except for the absence of R-parity.

In fact, since in this model the lightest $\hpi$ with the $\overline{\Psi}_{2}\Psi_{\bf 1}$ quantum number has a neutral component, one could imagine
obtaining a WIMP DM candidate as above by having an unbroken $\Psi_{\bf 1}$ species number. However, DM candidates with nonzero hypercharge are strongly
constrained by direct detection experiments and at mass scales where this WIMP would have the correct relic abundance, it is excluded by experiment.

Notice also that the quantum numbers of the hyperquarks in this model is
consistent with grand unification. The unification of the gauge
couplings can be better than that in the SM due to larger threshold
corrections enhanced by the hypercolor multiplicity factor $N$.

Finally, one can write down hyperbaryon number violating operators as in
(\ref{eq:HBviolating2}) and (\ref{eq:HBviolating3}), for instance in the
case of $N_{\rm HC}=3$ the operator
\begin{equation}
{\mathcal O}_{\rm HB}=\epsilon^{ABC}\left(\Psi^{\rm T}_{{\bf 3}A}\,{\mathcal C}\,\Psi_{{\bf 3}B}\right)\left(\overline{e}_{\rm R}\Psi_{{\bf
3}C}\right)+{\mathrm h.c.}
\end{equation}
can be added to the Lagrangian, where $e_{\rm R}$ is the right-handed electron field in the SM and the $\SU(3)_{\rm C}$ indices are contracted with an
$\epsilon^{ijk}$. For different choices of $N_{\rm HC}$, one can add insertions of $\Psi_{\bf 1}$, $\Psi_{\bf 2}$ or SM fields to make similar operators
gauge invariant.

%%%%%%%%%%%%%%%%%%%%%%%%%%%%%%%%
%%%%%%%%%%%%%%%%%%%%%%%%%%%%%%%%
\section{Summary and Conclusions}
\label{sec:conclusions}

We have studied the phenomenology of ``Vectorlike Confinement'', a broad
class of simple confining gauge theories with the confining scales
around a TeV. The new matter fields charged under the new gauge group
are assumed to be non-chiral under all SM gauge groups, which ensures
negligible impact on electroweak precision observables. The quantum
numbers of the new matter fields are such that their only renormalizable
couplings to the SM are gauge interactions. This guarantees that the
models are safe from precision flavor constraints, up to the effect of
nonrenormalizable operators which are suppressed by large mass scales.

At hadron colliders, spin-1 bound states of this new confining sector
can be resonantly produced via a mechanism exactly analogous to the
$\ga$-$\rho$ mixing in the resonant production of $\rho$ mesons in
$e^+$-$e^-$ collisions. These new spin-1 bound states can therefore be
regarded as a heavy gluon, a heavy $W^\pm$ or $Z$, depending on their
quantum numbers.

Instead of decaying back to SM particles, these spin-1 resonances
dominantly decay to a pair of lighter scalar bound states, analogously
to the dominance of the $\rho$ width by $\pi\pi$, which makes the models
insensitive to stringent resonance searches in dijets and dileptons.
Being pseudo-Goldstone bosons, these scalar bound states are naturally
light, in a few hundred GeV range.

These light scalar bound states exhibit a rich collider phenomenology.
They can be classified into two types. The scalars of the first type are
analogous to $\pi^0$ in the sense that they are made of a particle and
an antiparticle of the same species and can decay promptly to a pair of
gauge bosons. The scalars of the second type are analogous to $\pi^\pm$
in the sense that they are made of constituents of different species and
their decays require interactions that can change species. Just as
$\pi^\pm$ lives much longer than $\pi^0$ because changing species (from
$u$ to $d$) can only occur via nonrenormalizable 4-fermion operators,
our scalars of the second type are naturally long-lived.

Therefore, the generic collider signatures of Vectorlike Confinement are
the production of four gauge bosons from the decays of a pair of
``$\pi^0$-type'' scalars, and the production of a pair of
``$\pi^\pm$-type'' scalars. The latter is especially interesting
because, being made of different species, they typically carry
non-trivial SM charges, thereby appearing as CHAMPs or R-hadrons in
colliders. An observable fraction of these CHAMPs and R-hadrons are
produced from the decays of the aforementioned spin-1 resonances, so the
invariant mass distributions of CHAMP pairs and R-hadron pairs exhibit a
significant feature hinting at the existence of the parent spin-1
resonance.

In cases where an extra flavor structure in the coefficients of
\eq{L-pi-f-f} allows the mass scale suppressing these operators to be
lowered, the lifetimes of ``$\pi^\pm$-type'' scalars can be short enough
to decay inside the collider. In the models we discussed, they
preferentially decay to the 3rd generation, analogous to the preference
of the $\pi^\pm$ decay over muons to electrons. Therefore, depending on
their quantum numbers, they can appear as ``di-$\tau$,'' ``di-$b$,''
``di-top,'' or ``3rd-generation leptoquark.''

We have also discussed the possibility of a spectator species, which
occurs in an exotic SM representation and therefore is naturally
long-lived even on cosmological time scales, i.e.\ a dark matter
candidate. Heavier states carrying the spectator can cascade decay to
lighter ones, which can occur promptly or just slow enough to be visible
in the detector. Other models can exhibit SUSY-like spectra of scalars,
SUSY-like collider signatures (leptons + jets + missing $E_\text{T}$)
and dark matter candidates, as well as the possibility of accommodating
grand unification.

To conclude, a broad class of very simple gauge theories can exhibit
extremely rich and exotic collider phenomenology while having virtually
no constraints from existing precision measurements. In this sense the
spirit is very similar to those of ``Quirks'' \cite{quirks} and ``Hidden
Valley'' models \cite{hidden}. Taken altogether, it appears that there
is an excellent opportunity for the LHC to discover and explore a new
gauge symmetry and its dynamics beyond the SM.

\subsection*{Acknowledgments}

The authors would like to thank Richard Hill, David E. Kaplan, Kirill Melnikov, Matt
Strassler, Brock Tweedie and Jay Wacker for useful discussions and
insights. C.K. would also like to thank Steffen Schumann for assistance
in using Sherpa. The authors are supported by the National Science
Foundation grant NSF-PHY-0401513 and by the Johns Hopkins Theoretical
Interdisciplinary Physics and Astrophysics Center. C.K.~and T.O.~are
further supported in part by DOE grant DE-FG02-03ER4127 and by the
Alfred P.~Sloan Foundation. T.O.~is also supported by the Maryland
Center for Fundamental Physics.

%%%%%%%%%%%%%%%%%%%%%%%%%%%%%%%%
%%%%%%%%%%%%%%%%%%%%%%%%%%%%%%%%
\section*{Appendices}

\renewcommand{\thesection}{\Alph{section}}
\setcounter{section}{0}
\setcounter{subsection}{0}
\setcounter{subsubsection}{0}

%%%%%%%%%%%%%%%%%%%%%%%%%%%%%%%%
\section{A 2-flavor Model}
\label{app:2-flavor-model}

In this section we wish to study in detail one of the simplest models of
Vectorlike Confinement and illustrate some intriguing theoretical
subtleties that it contains. We take only one species of $\Psi$ which is
an SU(2) doublet:
\begin{center}
\begin{tabular}{c|c||c|c|c}
  & $\SU(N)_{\rm HC}$ & $\SU(3)_{\rm C}$ & $\SU(2)_{\rm W}$ & $\U(1)_{\rm Y}$  \\
\hline $\Psi$ & $\Box$ & ${\bf 1}$ & $\Box$ & 0
\end{tabular}
\end{center}
Note that $\Psi$ can have no renormalizable couplings to the Higgs or SM
fermions, rendering the model robustly safe from precision electroweak
and flavor constraints.

The $\hrho$ and $\hpi$ quantum numbers are
\beq
  (W^{\prime 0}, W^{\prime\pm}) &\sim& ({\bf 1}, {\bf 3})_0  \,,\nn\\
  \hrho^{\prime 0}           &\sim& ({\bf 1}, {\bf 1})_0  \,,\nn\\
  (\hpi^0, \hpi^\pm)      &\sim& ({\bf 1}, {\bf 3})_0  \,.
\eeq
The $\hpi$ mass from $W$ loops before EWSB is given by \eq{pi-mass-gauge}:
\beq
  m_{\hpi^0, \hpi^\pm} \simeq 0.13c \, m_\hrho  \,.
\eeq
After EWSB, the $\hpi^\pm$ and $\hpi^0$ masses split as in
\eq{mass-split-IR}, and the $\hpi^+$ will dominantly decay to
$W^{+*}\pi^0$. As in the case of \Sec{di-Champ}, the decay product of
this off-shell $W^+$ will be too soft to be detectable.

Now, naively, one might expect that $\hpi^0$ would decay promptly to
weak gauge bosons via the vertex \eq{pi-F-F}. This vertex, however,
vanishes for $\SU(2)$, and in fact there is a $\Z_2$ symmetry which
prevents a $\hpi^0$ from decaying. This $\Z_2$ symmetry is an accidental
symmetry of the renormalizable Lagrangian under which the hypercolor
gauge field $H_\mu$ and the hyperquarks $\psi$ and $\psi^\conj$
transform as
\beq
  \Psi   &\too& i\tau_2\,{\mathcal C}\,\Psi^{*}  \,,\nn\\
  H_\mu  &\too& -H_\mu^\T  \,,
\eql{Z_2} \eeq
where the Pauli matrix $\tau_2$ acts on the $\SU(2)_{\rm W}$ indices and
${\mathcal C}$ is the charge conjugation matrix. All SM fields are
invariant under this $\Z_2$. The $\lan\wba{\Psi}\Psi\ran$
condensate respects the symmetry upon confinement, so the $\Z_2$ is not
spontaneously broken. Note that a $\Z_2$ symmetry of this kind exists if and only if all hyperquarks are in (pseudo)real SM gauge representations (zero hypercharge and e.g. an adjoint of $\SU(3)_{\rm C}$). Such $\Z_2$ symmetries are studied extensively in \cite{Bai:2010qg}.

Now, note that the $\hpi$ is {\it odd} under the $\Z_2$ \eq{Z_2},
because the hyper-axial current
\beq
  J_{5}^{a\mu} \equiv \overline{\Psi}\fr{\tau^a}{2}\gamma_{5}\gamma^{\mu}\Psi\qquad (a=1,2,3)
  \eeq
is odd. Therefore, in the absence of nonrenormalizable terms in the
Lagrangian, the $\Z_2$ renders the $\hpi^0$ stable.

The $\Z_2$ symmetry is sensitive to non-renormalizable operators, but it
is interesting to contemplate whether it can be imposed exactly as the
$\hpi^0$ can be a good dark matter candidate. The problem is that the
fate of the $\hpi^0$ and the hyperbaryons are connected. Note that in
this model the hyperbaryons will have half-integer electric charge if
the number of hypercolor $N$ is odd, while they will be SM singlets if
$N$ is even. The lightest hyperbaryon in the former case would be
absolutely stable due to its fractional electric charge, which is
clearly excluded by observation. In the latter case, the choice of
$\Lambda_{HC}$ for which $\hpi^{0}$ has the correct relic density also
leads to the lightest hyperbaryon to have roughly the correct relic
density. Without performing a more sophisticated analysis, it is not
clear which will be the dominant dark matter component, therefore
drawing quantitative conclusions goes beyond the scope of this paper. On
the other hand, attempting to break hyperbaryon number generically also
breaks the $\Z_2$ symmetry, thereby rendering the $\hpi^{0}$ unstable,
such that one loses both dark matter candidates rather than only one of
them.

With the $\Z_2$ broken, the $\hpi^0$ will either decay to gauge boson
pairs or SM fermions, depending on which type of nonrenormalizable
operator dominates. The phenomenology of such decays is discussed in the
model presented in \Sec{di-Champ} where they happen generically, while
in this model they would only occur through additional $\Z_2$-violating
terms. For this reason we choose not to elaborate on the phenomenology
of this model as a representative case.

%%%%%%%%%%%%%%%%%%%%%%%%%%%%%%%%
%%%%%%%%%%%%%%%%%%%%%%%%%%%%%%%%
\section{Collider Constraints on Axion-like $\hpi$}
\label{app:axion-collider-const}

If an axion has a coupling to gluons as well as photons, there is a
potential constraint on $\hpi'_0$ production at hadron colliders with
subsequent decays to a pair of photons. We use the measurement presented
in \cite{Acosta:2004sn} to check whether the predicted diphoton cross
section is consistent with experiment. The simplest production mechanism
is $gg\rightarrow \hpi'_0$; while the production rate is loop suppressed
due to \eq{pi-F-F} and the branching fraction $\hpi'_0\rightarrow
\gamma\gamma$ is suppressed by $(\alpha_\text{EM} / \alpha_\text{s})^2$,
the gluon PDF's at low energy are large enough to give a sizeable cross
section for this process.

However, when a $\hpi'_0$ is produced this way, it has zero transverse
momentum, and the photons from the decay will not satisfy the
$p_\text{T}$ cuts used in \cite{Acosta:2004sn} unless $m_{\hpi'_0}\gsim
30$ GeV. Of course, this is an oversimplification, since there will
always be some amount of gluon radiation in the initial state, which the
axion can recoil against. In the case that this radiation is hard, one
can think of it as the $2\rightarrow 2$ process illustrated in figure
\fig{gg2ga}. However, the amplitude for this process has a singularity
and therefore cannot be used for the region of soft radiation. Thus, in
order to study whether \cite{Acosta:2004sn} rules out the existence of a
$\hpi'_0$, one needs to analyze a matched sample of $gg\rightarrow
\hpi'_0 + \text{jet(s)}$ events, which interpolates between the
$2\rightarrow 1$ approximation for heavy axions and the $2\rightarrow 2$
approximation for light axions.

To accomplish this, we have generated a matched event sample using
SHERPA 1.1 \cite{Gleisberg:2008ta} with ISR turned on. Implementing the
experimental cuts used in \cite{Acosta:2004sn}, we find that the
existence of $\hpi'_0$ is consistent with data for the reference values
of $f_{\hpi}=200$ GeV and the number of hypercolor $N=3$, for all
$m_{\hpi'_0}$.

While there are also contributions to axion production from
$qg\rightarrow \hpi'_0 + \text{jet(s)}$ and from $q\bar{q}\rightarrow
\hpi'_0 + \text{jet(s)}$, we find that these give cross sections much
smaller than $gg\rightarrow \hpi'_0 + \text{jet(s)}$.

Therefore, we conclude that there are no relevant constraints from
hadron colliders on the existence of axions in our model heavier than
100~MeV.

\begin{figure*}[t]
\begin{equation*}
  \lower6ex\hbox{\includegraphics[width=0.2\linewidth]{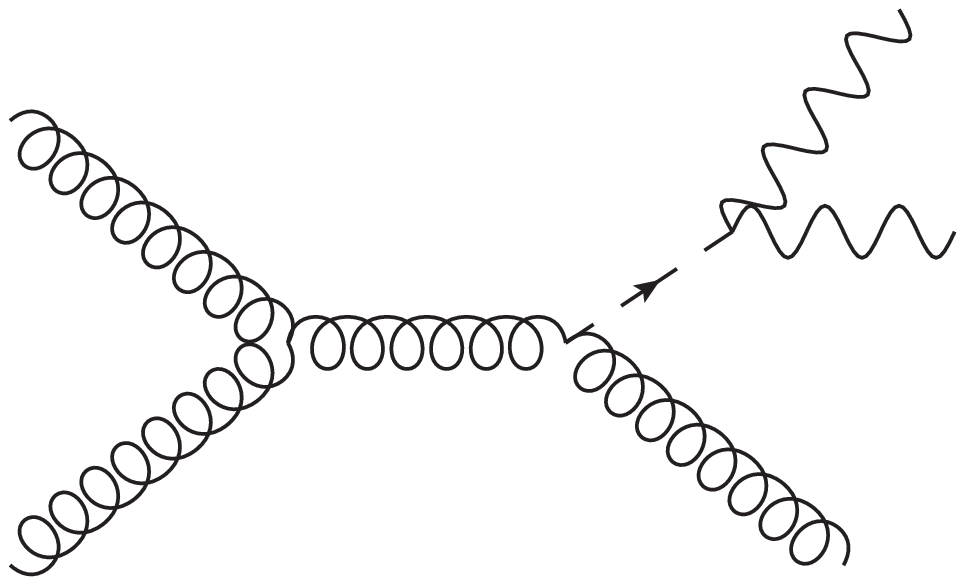}}
 +\lower13ex\hbox{\includegraphics[width=0.2\linewidth]{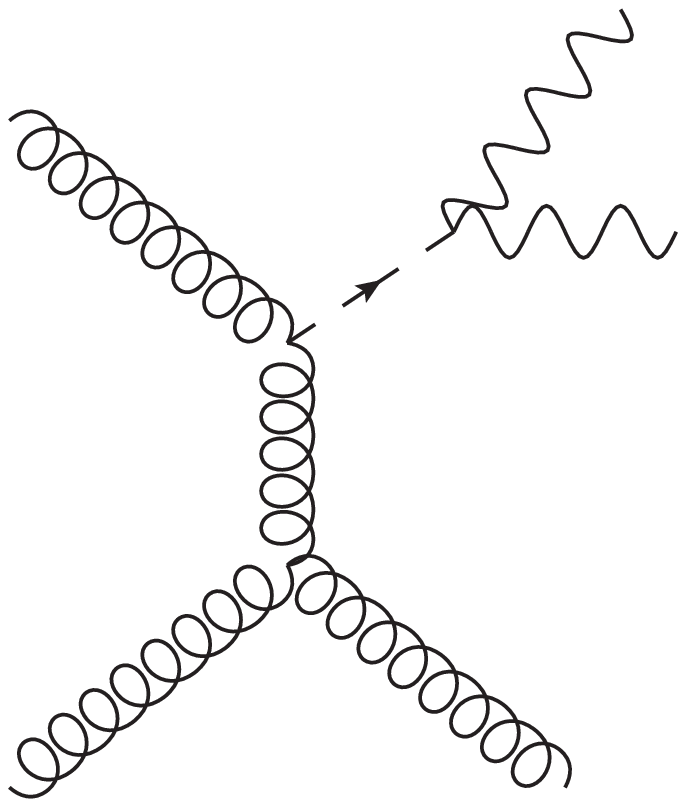}}
 +\lower11ex\hbox{\includegraphics[width=0.2\linewidth]{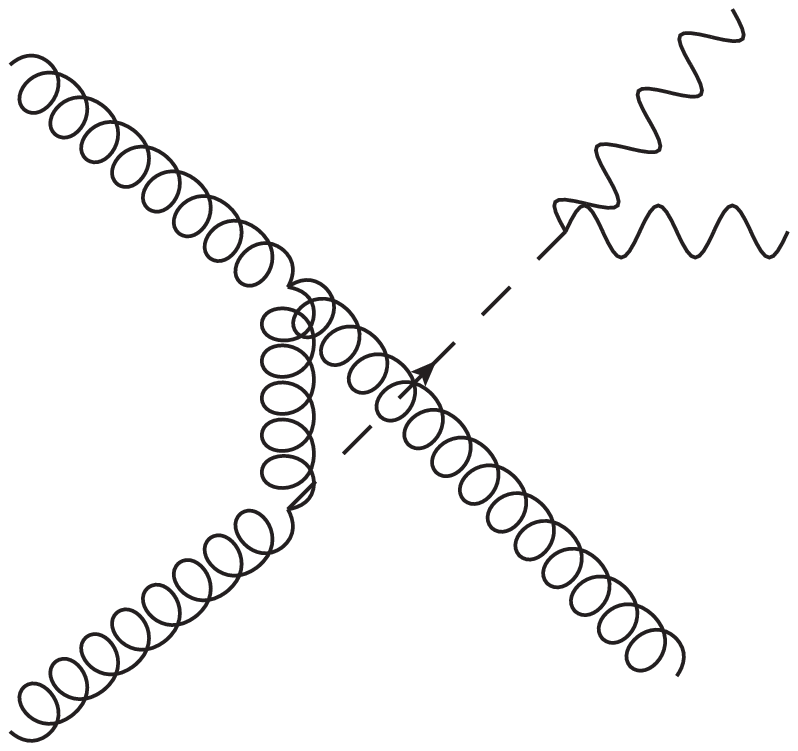}}
 +\lower9ex\hbox{\includegraphics[width=0.2\linewidth]{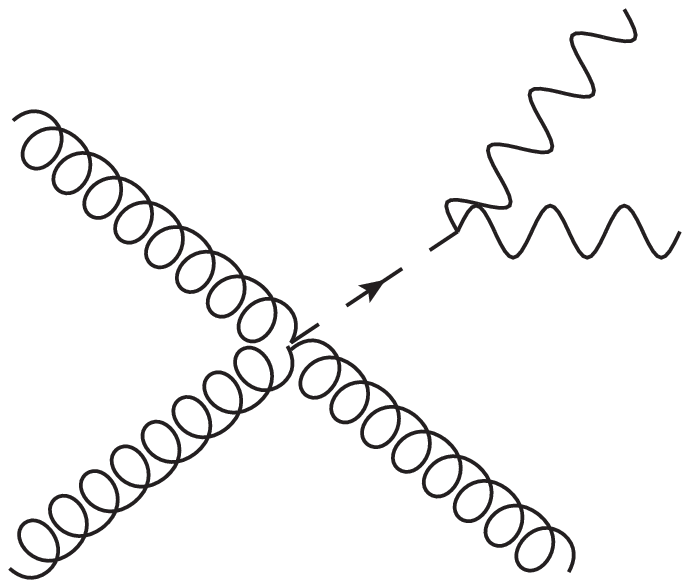}}
\end{equation*}
\caption{The Feynman diagrams contributing to the $gg\rightarrow\hpi'_0 g$ process.}
\figl{gg2ga}
\end{figure*}
%

%%%%%%%%%%%%%%%%%%%%%%%%%%%%%%%%
%%%%%%%%%%%%%%%%%%%%%%%%%%%%%%%%

\end{document}